\documentclass[aps,prb,eqsecnum]{revtex4}
\newcommand{\diag}{{\rm diag\,}}
\newcommand{\vA}{{\bf A}}
\newcommand{\vk}{{\bf k}}
\newcommand{\vecr}{{\bf r}}
\newcommand{\EM}{{\rm EM}}
\newcommand{\diff}{{\rm d}}
\newcommand{\e}{{\rm e}}
\begin{document}
\title{Quantum electrodynamic fluctuations
       of the macroscopic Josephson phase}

\author{H.~Kohler$^1$, F.~Guinea$^2$, and F.~Sols$^1$}

\affiliation{$^1$ Departamento de F\'{\i}sica Te\'orica de la Materia
Condensada e Instituto Nicol\'as Cabrera, Universidad
                  Aut\'onoma, E-28049 Madrid, Spain\\
              $^2$ Instituto de Ciencia de Materiales, CSIC,
              Campus de Cantoblanco, E-28049 Madrid, Spain}
\date{\today}
\begin{abstract}
 We study  the equilibrium dynamics of the relative phase in a
superconducting Josephson link taking into account the quantum
fluctuations of the electromagnetic vacuum. The photons act as a
superohmic heat bath on the relative Cooper pair number and thus,
indirectly, on the macroscopic phase difference $\phi$. This leads to
an enhancement of the mean square $\langle\phi^2\rangle$ that adds to
the spread due to the Coulomb interaction carried by the longitudinal
electromagnetic field. We also include the interaction with the
electronic degrees of freedom due to quasiparticle tunneling, which
couple to the phase and only indirectly to the particle number. The
simultaneous inclusion of both the radiation field fluctuations and
quasiparticle tunneling leads to a novel type of particle--bath
Hamiltonian in which the quantum particle couples through its position
and momentum to two independent bosonic heat baths. We study the
interplay between the two mechanisms in the present context and find
interference contributions to the quantum fluctuations of the phase. We
explore the observability of the QED effects discussed here.
\end{abstract}
\maketitle
\section{Introduction}
\label{Intro}A proper understanding of decoherence as resulting from
the dissipative effect on a quantum system of a complex environment is
of paramount importance in diverse areas of modern physics and
technology. The influence of many degrees of freedom on a given quantum
variable has for long been considered a problem of fundamental interest
\cite{feyn63}, partly because of its connection to the polaron problem
\cite{feyn55}. However, it was only in the eighties when, because of
its relevance to macroscopic quantum tunneling and coherence
 \cite{cald83,legg87}, the problem was addressed by a large community
of physicists. The loss of quantum wave coherence -also referred to as
decoherence or dephasing- appears in a wide range of physical contexts.
It has been generally recognized that decoherence acts much faster than
energy dissipation, since it only requires the excitation or
destruction of a single quantum of the thermal oscillator bath, while
many quanta are necessary to change the particle energy appreciably
\cite{unru89,ster90,zapa03}. In the context of mesoscopic electron
systems, dephasing is responsible for the destruction of those quantum
interference effects which characterize transport through ballistic
nanostructures or disordered metals at low temperatures. Particle-bath
models have been used to address such a problem \cite{ster90,sols92}.
In disordered metals at sufficiently low temperatures, the dominant
mechanism responsible for the loss of phase coherence of electrons near
the Fermi surface is the Coulomb interaction with identical electrons
in the metal \cite{alts82}. The question of whether or not the electron
dephasing time diverges at zero temperature has been the subject of a
recent controversy \cite{controversy}.

We wish to emphasize here that, under the generic term decoherence or
dephasing, one may be contemplating physical problems of widely
different conceptual and technical nature which may yield different
answers to apparently similar questions. For instance, the general
question `does zero-point decoherence exist?' does not admit a
universally valid response. The answer depends on the system under
attention, the nature of its surrounding environment, the properties of
the coupling, and the physical representation within which we frame the
question. The latter remark refers to the importance of the basis set
in whose representation the reduced density matrix of the quantum
system tends to become diagonal as a result of the interaction with the
environment. The choice of representation is related in turn to the
observable to be measured, since the expectation value must be
sensitive to the off-diagonal terms of the density matrix in the
desired representation. As examples of distinct physical problems
having to do with zero-point fluctuations, we may mention the problem
of decoherence for a quantum particle coupled to a dissipative bath,
which is qualitatively different from that posed by Fermi surface
electrons interacting with other electrons in a disordered metal. An
intermediate class of problems is that of conduction electrons
interacting with an external environment, which has been shown to
modify some Fermi liquid properties such as persistent currents in
mesoscopic rings \cite{cedr01}.
The message is that possible answers to questions having to do with the
loss of quantum wave coherence must be investigated case by case and
cannot be translated from one physical context to another without a
careful scrutiny.

The general problem of decoherence is receiving renewed attention due
to its central role in the design of quantum information processing
systems. In this context, it seems that all potential sources of
decoherence must be explored. Due to its ubiquity, the quantum
electrodynamic (QED) field provides the most basic decoherence
mechanism that a charged particle may experience. The purpose of this
paper is to investigate the role of the quantum electromagnetic (EM)
field as a possible source of decoherence. Specifically, we aim at
understanding the effect of the {\em transverse} component of the EM
field. It must be distinguished from the {\em longitudinal} component,
which in the radiation gauge is originated by the electrostatic Coulomb
interaction among charged particles \cite{jack75}. The combined effect
of the transverse and the longitudinal field in a many-body context has
been recognized to be responsible for logarithmic departures from the
Fermi liquid picture in two-dimensional electron systems \cite{two-dim}
and for the reduction of persistent currents in mesoscopic rings
\cite{loss93}. In the context of particle-bath problems, it is fair to
say however that, despite its fundamental character, the QED field has
been little explored as a dissipative environment that may disturb
electrons and, in general, charged particles. From the quantum
dissipation viewpoint, Quantum Electrodynamics poses at least two new
problems. First, in vacuum it is a superohmic environment
\cite{baro91,zapa97} and, as such, it has been much less studied than
its ohmic counterparts. Second, it has no intrinsic upper cutoff. Thus,
when needed, appropriate cutoffs must be introduced on the basis of
sound physical arguments.

Some studies can be found in the literature dealing with the coherence
properties of free charged particles in the presence of the background
radiation field \cite{baro91,joos85,ford93,haba01}. To understand in
greater depth the effect of the transverse photon field as a source of
dissipation, we choose as our case study a system which is otherwise
quite well understood: the phase-number variable of a Josephson link
connecting two superconductors. Typically, such a weak link is achieved
with a tunneling barrier or a point contact. The problem which we wish
to address here may be viewed as the {\em macroscopic}
quantum-mechanical version of the {\em Lamb shift} problem, since we
aim at calculating the effect of the fluctuations of the transverse
photon field on an otherwise conservative quantum-mechanical problem.
In particular we will focus on the calculation of the uncertainty
$\Delta \phi \equiv \langle\phi^2\rangle - \langle\phi\rangle^2$, where
$\phi$ is the relative phase between the two superconductors. We will
focus on the equilibrium dynamics, for which $\langle\phi\rangle=0$,
and on the harmonic limit $\langle\phi^2\rangle\ll 1$. The effect of
the transverse EM field on $\langle\phi^2\rangle$ can be potentially
detected, for instance, as a Debye-Waller reduction of the Josephson
critical current or as an intrinsic source of uncertainty in the fine
measurement of tiny magnetic fields.

This paper builds on a preliminary study presented in Ref.
 \cite{zapa97}, where it was shown that, in the absence of a
Josephson coupling, recurrent dynamics alternating between quantum
collapses and revivals is robust against fluctuations of the QED field
\cite{sols94}. Such collapses and revivals were later predicted
\cite{imam97} and observed \cite{grei02} in atomic Bose-Einstein
condensates. However, a number of theoretical questions, including the
role of the upper frequency cutoff, were left unexplored. In the
present work, we also investigate the effect of quasiparticles, which
cause fluctuations of the longitudinal field and which we approximate
by effective oscillators, adapting the work of Ref. \cite{ecke82} to a
Hamiltonian description. When computing the combined effect of both the
QED and quasiparticle fields, we find that, in the denominators of the
spectral weight contributing to $\langle\phi^2\rangle$, there are terms
which are caused by the interference between the two baths.

In Ref. \cite{rogo74} current and voltage fluctuations in tunnel
junctions were studied. There special attention was paid to the biased
case and to the effect of fluctuations on the $I-V$ characteristics.
Here we have focussed on the calculation of the equilibrium values of
$\langle\phi^2\rangle$ and $\langle N^2\rangle$ and have used a
consistent particle-bath approach. Moreover, we have attempted to
clearly separate the transverse and longitudinal electromagnetic fields
and have modelled quasiparticles with effective oscillator baths. Our
work presents a compact and somewhat involved treatment of the effect
of the photon and quasiparticle fields on the phase of a
superconducting weak link. Fortunately, some aspects of this complex
problem may be understood in simpler terms for some particular cases.
This observation motivates section II, which is devoted to provide a
largely self-contained preview of some of the results to be derived
more rigorously later. That discussion is inspired to some extent in
simplified derivations of the atomic Lamb shift (see e.g. Ref.
\cite{bjor64}). Starting from the classical Langevin equation satisfied
by $\phi$, the EM field is introduced by invoking gauge invariance
arguments. In the simplest cases, we are able to derive fourth-order
differential equations which, in the absence of quasiparticles, reduce
to Abraham-Lorentz equations for both $N$ and $\phi$. We argue
conclusively that the coupling to the EM field is essentially that of a
{\em fluctuating dipole}.

Section III deals in detail with the coupling of the macroscopic
superconducting phase to the QED field, which is obtained invoking
gauge invariance and introducing appropriate canonical transformations.
The number and phase autocorrelation functions are calculated. This
permits to compute the QED correction to the longitudinal field result
\cite{scho90} for $\langle\phi^2\rangle$ in terms of the junction
aspect ratio and the {\em fine structure constant}. In section IV, we
study the coupling to the quasiparticle bath, i.e. to the many-body
fluctuations of the longitudinal field. We adapt the work of Ref.
\cite{ecke82} to a Hamiltonian language where the dynamics is
investigated through the Heisenberg equations of motion. Section V
addresses the most complex issue, namely, the combined effect of the
photon and quasiparticle fields. There we face a fundamental problem in
quantum dissipation that has so far received little attention: the
behavior of a quantum particle interacting with two different baths
through its position and momentum \cite{legg84}. Finally, section VI is
devoted to a concluding discussion.

\section{General remarks on the equations of motion}
\label{GEQ} The standard way to describe the dynamics of the
macroscopic phase in Josephson links is the resistively and
capacitively shunted junction (RCSJ) model
\cite{scho90,baro82,tink96,kett99}, which contemplates an ideal
Josephson junction shunted by a resistor and a capacitor. The resistor
models the dissipative effect of incoherent quasiparticle tunneling
through the junction, while the capacitor accounts for the charging
energy, which plays the role of a kinetic energy for the phase. In the
absence of driving currents, the RSCJ model reads
\begin{eqnarray}
\dot{N}(t)&=&-\frac{E_J}{\hbar} \sin \phi(t)
-\frac{\hbar}{4e^2R}\dot{\phi}(t)+ \frac{1}{2e}I(t) \label{int0}
\\
\dot{\phi}(t)&=& \frac{E_C}{\hbar}N(t), \label{int1}
\end{eqnarray}
where $N$, the number of transferred Cooper pairs across the junction,
is canonically conjugate to the relative phase $\phi$. We have
introduced the notation $E_J\equiv\hbar I_c/2e$ and $E_C\equiv 4e^2/C$
for the Josephson coupling energy and the charging energy respectively
($I_c$ and $C$ are the critical current and the capacitance of the
junction). $I(t)$ is a stochastic process with zero mean. At high
temperatures it is related to the resistance by Einstein's relation
$\langle I(t)I(0)\rangle=2k_BT\delta(t)/R$, where $R$ is the resistance
of the junction in the normal state. Eq.~(\ref{int1}) is recognized as
the ac Josephson relation written in terms of the relative Cooper pair
number, which generates a chemical potential difference $\mu$ through
the interaction energy $E_C=\partial \mu/\partial N$.

The effect of the EM vacuum fluctuations may be introduced through the
following argumentation: Written in language of the Coulomb gauge (as
is standard in these physical contexts \cite{baro82,tink96}), Eqs.
(\ref{int0}) and (\ref{int1}) relate the phase to gauge invariant
quantities. More specifically, they include the effect of the
longitudinal electric field, which in its simplest form yields a
circulation $\int_1^2{\bf E}_\parallel\cdot \diff\vecr=2eN/C$.
Therefore, we are entitled to replace the phase in (\ref{int1}) by its
gauge invariant expression $\phi=\phi_1-\phi_2+(2e/\hbar c)\int_1^2
\diff\vecr\cdot {\bf A}({\bf r})$, where 1 and 2 are points deep enough
in the bulk of superconductors 1 (left) and 2 (right). The resulting
equation is again interpreted in the particular language of the
transverse gauge, in which the vector potential is characteristically
related to the transverse electric field via $c{\bf E}_\perp=-\dot{{\bf
A}}$. This allows us to write Eq.~(\ref{int1}) in a form which treats
the transverse and longitudinal electronic field on the same footing:
\begin{equation}
\dot{\phi}= \frac{2e}{\hbar}\left(\int_1^2{\bf E}_\parallel\cdot
            \diff\vecr+\int_1^2{\bf E}_\perp\cdot \diff\vecr\right)\ .
            \label{int1a}
\end{equation}
Thus one may view the transverse EM modes as the cause of additional
voltage fluctuations $V(t)=\int_1^2{\bf E}_\perp\cdot \diff\vecr$ with
zero mean $\langle V(t)\rangle=0$. This interpretation is physically
appealing but still insufficient to understand in depth the detailed
nature of the effect of ${\bf E}_\perp$ on $\phi$.

    For one thing, we may wonder whether the fluctuating transverse
voltage may generate a ``slow'' contribution to the r.h.s. of
(\ref{int1}) which would relate to the ``fast'' part through an
appropriate expression of the fluctuation-dissipation theorem. This
would be analogous to the already noted relation between the second and
third terms on the r.h.s. of Eq. (\ref{int0}). In this regard we note
that the slow term proportional to $\dot{\phi}$ in (\ref{int0}) is but
the Markovian limit of a more general retarded expression
\cite{weis99},
\begin{equation} \label{markovian}
\int_{-\infty}^t \Gamma_{\rm qp}(t-t^\prime)\dot{\phi}(t^\prime)\diff
t^\prime \longrightarrow (\hbar/4e^2R)\dot{\phi}(t)\ .
\end{equation}
Clearly, the fluctuating current $I(t)$ plays a role for $\phi$
analogous to that which the fluctuating transverse potential $V(t)$
represents for $N$. This observation suggests that $N$ and $\phi$
satisfy formally similar retarded equations of motion, which must be
the form
\begin{eqnarray}
\dot{N}(t)&=&  -\frac{E_J}{\hbar}\sin\phi(t)-\int_{-\infty}^t
      \Gamma_{\rm qp}(t-t^\prime)\dot{\phi}(t^\prime)\diff t^\prime
               +\frac{1}{2e}I(t) \label{int2-0}\\
\dot{\phi}(t)&=& \frac{E_C}{\hbar} N(t) +
              \int_{-\infty}^t
          \Gamma_{\rm EM}(t-t^\prime)\dot{N}(t^\prime)\diff t^\prime+
              \frac{2e}{\hbar} V(t)\ . \label{int2}
\end{eqnarray}
A rigorous derivation and ultimate justification of this result is
possible after the analysis given in Secs.~\ref{SECEM1}, \ref{HB}, and
\ref{simult}, where particle-bath couplings are derived which yield
Eqs. (\ref{int2-0}) and (\ref{int2}) as the Heisenberg equations of
motion for $N$ and $\phi$. The dissipation kernels $\Gamma_{\rm EM}(t)$
and $\Gamma_{\rm qp}(t)$ are given in terms of the spectral functions
$J_{\rm EM}(\omega)$ and  $J_{\rm qp}(\omega)$, respectively, through
the relations
\begin{eqnarray}
\Gamma_{\rm EM}(t)=\int_0^\infty J_{\rm EM}(\omega)\cos(\omega t)
\diff\ln(\omega) \label{gj-1}\\ \Gamma_{\rm qp}(t)=\int_0^\infty J_{\rm
qp}(\omega)\cos(\omega t)\diff\ln(\omega)\ .\label{gj-2}
\end{eqnarray}
The current and voltage fluctuations are related to the spectral
functions through the fluctuation-dissipation theorem,
\begin{eqnarray}
\langle V(t)V(0)\rangle &=&
                      \frac{\hbar^2}{8e^2}
                      \int_{-\infty}^\infty J_{\rm EM}(\omega)
                      \frac{\exp(i\omega t)}
                      {1-\exp\left(-\hbar\beta\omega\right)}
                      \diff\omega
                      \label{int3a-0}\\
\langle I(t)I(0)\rangle &=& 2e^2
                     \int_{-\infty}^\infty J_{\rm qp}(\omega)
                     \frac{\exp(i\omega t)}
                     {1-\exp\left(-\hbar\beta\omega\right)}
                     \diff\omega\ . \label{int3a}
\end{eqnarray}
As expressed in Eqs.~(\ref{int2-0}) and (\ref{int2}), the symmetry
between $\phi$ and $I(t)$ on the one hand and $N$ and $V(t)$ on the
other hand, is most evident. Both stochastic ``sources" obey two
independent fluctuation--dissipation theorems. The study of this type
of ``mixed" dissipation has been pioneered by Leggett \cite{legg84} for
quantum systems. Dissipation due to coupling to the momentum variable
has been addressed for an ohmic bath in Ref.~ \cite{cucc01}. The formal
symmetry between $\phi$ and $N$ is broken not only by the different
potential terms, already present in Eqs.~(\ref{int2-0}) and
(\ref{int2}), but also by the different form of the spectral functions
$J_{\rm EM}(\omega)$ and $J_{\rm qp}(\omega)$ describing the noise.
$J_{\rm qp}(\omega)$ is usually taken to be ohmic \cite{scho90,weis99}
\begin{equation}
J_{\rm qp}(\omega)= \frac{\hbar\omega}{2\pi e^2R},
\label{ohmic}\end{equation} which through Eq. (\ref{markovian}) makes
Eq.~(\ref{int2-0}) become Eq.~(\ref{int0}). It also guarantees that Eq.
(\ref{int3a}) achieves the correct high temperature limit in the form
of the Einstein relation mentioned above. By contrast, we expect that,
in three-dimensional free space, the spectral function of the
electromagnetic field is cubic. Actually, a detailed calculation
performed in Sec.~\ref{SECEM1} yields
\begin{equation} \label{superohmic}
J_{\rm EM} (\omega)= \frac{8d^2\alpha\omega^3}{3\pi c^2},
\end{equation}
where $d$ is the distance between the two electronic clouds and $\alpha
\simeq 1/137$ is the fine structure constant. Moreover, the spectral
function $J_{\rm EM}(\omega)$ has to be regularized by an upper cutoff
frequency $\omega_c$ which is the inverse response time of the
electromagnetic field. This point is discussed in detail in
Sec.~\ref{SECEM1}, where the cutoff frequency is shown to be determined
by the surface area of the electrodes. Typical values for the diameter
of the electrodes of Josephson point contacts are about $100$ nm. This
translates to a frequency of $\omega_c\simeq 2\pi\times 10^{18}$ Hz
which is higher than typical values of the Josephson plasmon frequency
$\omega_{\rm JP}=\sqrt{E_J E_C}=\sqrt{2eI_c/\hbar C}\sim 2\pi \times
10^{10}$ Hz by several orders of magnitude.

Being the cutoff frequency high, we may approximate Eq.~(\ref{int2}) by
a form which is local in time:
\begin{equation}
\dot{\phi}(t)=\frac{E_C}{\hbar}  N(t)+
                \frac{8\alpha d^2\omega_c}{3\pi c^2}\ddot{N}(t)
                 -\frac{4\alpha d^2}{3c^2}
                 \stackrel{\cdot\cdot\cdot}{N}(t)
                 +\frac{2e}{\hbar}V(t)\quad .
\label{int4}
\end{equation}
Eliminating $\phi$ in favor of $N$ and viceversa, Eqs. (\ref{int0}) and
(\ref{int4}) become two decoupled fourth-order differential equations
whose inhomogeneous, fast fluctuating parts are however related to each
other. Specifically, if we approximate $\sin \phi \simeq \phi$, we
obtain
\begin{eqnarray}
&&-\frac{\hbar\alpha d^2}{3e^2Rc^2}
\stackrel{\cdot\cdot\cdot\cdot}{N}(t)
  -\frac{4\alpha d^2}{3c^2} \left(\frac{E_J}{\hbar}-
   \frac{\hbar\omega_c}{2\pi e^2R}\right)
   \stackrel{\cdot\cdot\cdot}{N}(t)+
\left(1+\frac{8\alpha d^2E_J\omega_c}{3\pi\hbar c^2}\right)\ddot{N}(t)
+\frac{1}{RC}\dot{N}(t)+\omega_{\rm JP}^2N(t)=\nonumber\\
 &&\qquad\qquad\qquad\qquad\qquad\qquad\qquad\qquad\qquad
   \frac{1}{2e}\dot{I}(t)-
  \frac{2eE_J}{\hbar^2}V(t)-\frac{1}{2eR}\dot{V}(t)\ ,
  \label{int5-0}\\
  \nonumber \\
&&-\frac{\hbar\alpha d^2}{3e^2Rc^2}
\stackrel{\cdot\cdot\cdot\cdot}{\phi}(t) -\frac{4\alpha d^2}{3c^2}
\left(\frac{E_J}{\hbar}-\frac{\hbar\omega_c}{2\pi
e^2R}\right)\stackrel{\cdot\cdot\cdot}{\phi}(t)+\left(1+\frac{8\alpha
d^2E_J\omega_c}{3\pi \hbar
c^2}\right)\ddot{\phi}(t)+\frac{1}{RC}\dot{\phi}(t)+\omega_{\rm
JP}^2\phi(t)=\nonumber\\
&&\qquad\qquad\qquad\qquad\qquad\qquad\qquad   \frac{E_C}{2e\hbar}I(t)+
\frac{8\alpha d^2\omega_c}{3\pi c^2}\ddot{I}(t)-
  \frac{2\alpha d^2}{3ec^2}\stackrel{\cdot\cdot\cdot}{I}(t)+
  \frac{2e}{\hbar}\dot{V}(t)\ .
\label{int5}
\end{eqnarray}
Here we stress that $I(t)$ and $V(t)$ are two uncorrelated stochastic
processes. Nevertheless, their simultaneous coupling to the junction
generates interference terms. The most obvious signature of such
interference is the fourth order time derivative in Eqs.~(\ref{int5-0})
and (\ref{int5}), since that term vanishes whenever any of the two
coupling constants $R^{-1}$ or $\alpha$ vanishes.

We observe that the left hand sides of Eqs. (\ref{int5-0}) and
(\ref{int5}) are identical except for an interchange of $N$ and $\phi$.
This means that the stochastic differential equations satisfied by $N$
and $\phi$ only differ in the fast fluctuating part. The structure of
Eqs. (\ref{int5-0}) and (\ref{int5}) suggests the introduction of a
generalized ``response" function
\begin{eqnarray}
&&\chi^{-1}(\omega)=\omega_{\rm
JP}^2-\omega^2+\frac{E_J}{\hbar}\left(\frac{i\pi}{2}
        -\frac{\omega_c}{\omega}\right)J_{\rm EM}(\omega)+
         \frac{E_C}{\hbar}\frac{i\pi}{2}J_{\rm qp}(\omega)
+\frac{i\pi}{2}J_{\rm qp}(\omega)J_{\rm
EM}(\omega)\left(\frac{i\pi}{2}-
         \frac{\omega_c}{\omega}\right) \ .
\label{int6}
\end{eqnarray}
A more general form of the response function (\ref{int6}), valid for
arbitrary spectral densities, will be given in Sec.~\ref{CqpEM}. In
Fourier space, Eqs.~(\ref{int5-0}) and (\ref{int5}) read
\begin{eqnarray}
N(\omega)&=&\chi(\omega)\left[\frac{i\omega}{2e}I(\omega)-
           \frac{2e}{\hbar}\left(\frac{E_J}{\hbar}+
           \frac{i\pi}{2}J(\omega)\right)V(\omega)\right]
\label{int7-0} \\
\nonumber\\
\phi(\omega)&=&\chi(\omega)\left[\frac{i\omega2e}{\hbar}V(\omega)+
           \left(\frac{E_C}{\hbar}+\frac{i\pi}{2}J(\omega)
           -\frac{\omega_c}
           {\omega}J(\omega)\right)\frac{1}{2e}I(\omega)\right]\ .
\label{int7}
\end{eqnarray}
Ultimately we are interested in autocorrelation functions  of the form
\begin{equation}
\langle\phi(t)\phi(t^\prime)\rangle=
\frac{1}{2\pi}\int\langle|\phi(\omega)|^2\rangle
\e^{i\omega(t-t^\prime)} \diff\omega\ ,
 \label{int8}
\end{equation}
where interference terms must clearly play a role.

QED effects can be studied more clearly in Eqs.~(\ref{int5-0}) and
(\ref{int5}) if one neglects the effect of the electronic environment,
i.e. if one eliminates in Eqs.~(\ref{int5-0}) and (\ref{int5}) all
terms depending on $R^{-1}$ or $I(t)$. One obtains
\begin{eqnarray}
-\frac{4 E_J\alpha d^2}{3c^2\hbar}\stackrel{\cdot\cdot\cdot}{N}(t)+
             \left(1+\frac{8\alpha d^2E_J\omega_c}
             {3\pi \hbar c^2}\right)\ddot{N}(t)+
             \omega_{\rm JP}^2N(t)&=&-\frac{2eE_J}{\hbar^2}V(t)
             \label{int9-0}\\
-\frac{4 E_J\alpha d^2}{3c^2\hbar}\stackrel{\cdot\cdot\cdot}{\phi}+
             \left(1+\frac{8\alpha d^2E_J\omega_c}
             {2\pi \hbar c^2}\right)
             \ddot{\phi}(t)+
             \omega_{\rm JP}^2\phi(t)&=&\frac{2e}{\hbar} \dot{V}(t),
\label{int9}
\end{eqnarray}
which are equations of the Abraham--Lorentz type \cite{jack75}. The
fact that the EM field couples to the particle number and not
(directly) to the phase is reflected in the different r.h.s. of Eqs.
(\ref{int9-0}) and (\ref{int9}), which are proportional to $V(t)$ and
$\dot{V}(t)$, respectively. This means that, although their slow parts
obey identical equations, $\phi$ and $N$ experience different
fluctuations terms and thus possess different power spectra. Close
inspection reveals that (\ref{int9-0}) and (\ref{int9}) are the
classical equations of motion for a dipole in a harmonic potential
interacting with its own radiation. Indeed it could have been
anticipated that the equation of motion for $N$ should be that of a
radiating dipole, since the relative particle number $N$ couples to the
EM vacuum through the electric dipole which it generates, $P=2eNd$. In
effect, neglecting the fluctuating term as well as the mass
renormalization terms, Eq.~(\ref{int9-0}) can be written as
\begin{equation}
\ddot{P}(t)+\omega_{\rm
JP}^2P(t)-\tau\stackrel{\cdot\cdot\cdot}{P}(t)=0 \ ,\label{int10}
\end{equation}
with $\tau=4E_J\alpha d^2/3\hbar c^2$. For a radiating harmonic dipole
of charge $e$ and mass $m$ the characteristic time $\tau$ is given by
\cite{jack75} $\tau=2e^2/3mc^3$. We conclude that the equation of
motion for $N$ is that of a radiating dipole with effective mass
$m=\hbar^2/3E_Jd^2$.

\section{Coupling to the electromagnetic environment}
\label{SECEM1} A general description of the electrodynamics in a
Josephson junction with electrodes in the $x-y$ plane is given by a two
dimensional sine--Gordon Hamiltonian for the position dependent phase
\cite{baro82} $\phi(x,y)$. The characteristic length scale of the
theory is the Josephson penetration depth $\lambda_J$. A small
Josephson junction is characterized by an area $A$ of the electrodes
which is much smaller than $\lambda_J$. In this limit spatial
variations of the phase in the $x-y$ plane are energetically highly
unfavorable. They are suppressed like $\sim \sqrt{A}/\lambda_J$. Then,
the sine-Gordon model reduces to the Hamiltonian of a pendulum, first
discussed by Anderson \cite{ande64}
\begin{equation}
H_0=E_J\left(1-\cos\phi\right)+\frac{E_C}{2}N^2+H_{\rm EM}\ .
\label{ee0}
\end{equation}
The phase difference across the junction and the number of transferred
Cooper pairs are canonical conjugates: $[\phi,N]_-=i$. As indicated in
Sec.~\ref{GEQ}, we introduce the coupling to the electromagnetic
environment replacing $\phi$ by its gauge-invariant expression
$\phi=\phi_1-\phi_2+\Lambda_{12}$, where $\Lambda_{12}\equiv(2e/\hbar
c)\int_{1}^{2} \diff\vecr\cdot{\bf A}({\bf r})$. The line integral
connects two points $1$ and $2$ lying deep enough in the left and right
superconductor, where phase fluctuations are completely suppressed.
That means, it extends over a minimal length of $2\lambda_L$, where
$\lambda_L$ is the penetration depth of the magnetic field. In
dissipative quantum systems one usually assumes the influence of the
system onto the bath to be small. However in our case the system is
made by superconductors which of course alter drastically the low
frequency behavior of the electromagnetic vacuum. The screening of the
EM field out of the superconductor should be taken into account by an
appropriate choice of the wave functions $A_{\vk\lambda}({\bf r})$ of
the normal modes $\vk\lambda$ in terms of which the vector potential
$\vA(\vecr)$ is expanded. These must be the modes which diagonalise the
EM Hamiltonian. Then we can write
\begin{eqnarray}
{\bf A}({\bf r})&=&
           \sum_{\vk\lambda}
           \left(\frac{2\pi\hbar c^2}{\omega_k}\right)^{1/2}\;
           {\vec{\varepsilon}_\lambda}\left(A_{\vk\lambda}({\bf r})\;
                 a_{\vk\lambda}\;+\; {\rm H.c.}\right)\ ,\nonumber\\
H_{\EM}&=&\sum_{\vk\lambda}\hbar\omega_k a_{\vk\lambda}^\dagger
          a_{\vk\lambda}\ .
\label{ee1}
\end{eqnarray}
The influence of the superconductor on the EM field is encoded in the
set of orthonormal functions $A_{\vk\lambda}(\vecr)$. They are
solutions of a Helmholtz equation with boundary conditions imposed by
the geometry of the junction and the nature of the superconducting
state. $H_{\EM}$ is expressed in its usual form. It is also convenient
to expand
\begin{eqnarray}
\Lambda_{12}&=&\sum_{\vk\lambda}
      \left(\frac{8\pi e^2}{\hbar\omega_k}\right)^{1/2}
      \left( \Lambda_{\vk\lambda} a_{\vk\lambda}+
      \Lambda_{\vk\lambda}^*a_{\vk\lambda}^\dagger\right)\ ,
      \nonumber\\
\Lambda_{\vk\lambda}&=&\int_1^2
      A_{\vk\lambda}({\bf r}) {\vec{\varepsilon}_\lambda}  \cdot
      \diff\vecr
      \label{ee1a}
\end{eqnarray}
Shifting $\phi$ by $-\Lambda_{12}$ via a unitary transformation
$H\rightarrow U^{-1}HU$ with $U=\exp(iN\Lambda_{12})$ one obtains
\begin{equation}
H=H_0 -N\sum_{\vk\lambda}(8\pi e^2\hbar\omega_k)^{1/2}
\left(i\Lambda_{\vk\lambda} a_{\vk\lambda}
-i\Lambda^*_{\vk\lambda}a^\dagger_{\vk\lambda}\right)+8\pi e^2
N^2\sum_{\vk\lambda}|\Lambda_{\vk\lambda}|^2\ .
\label{ee4}\end{equation} If the superconductors are equal on both
sides one has a reflection symmetry with respect to the $x-y$ plane.
This symmetry and the trivial observation that
$\Lambda_{12}=-\Lambda_{21}$ are sufficient to conclude that all
$\Lambda_{\vk\lambda}$ are antisymmetric in $z_2-z_1$ and purely real.
This allows us to further simplify Eq.~(\ref{ee4}). Finally, with
$a_{\vk\lambda}\rightarrow ia_{\vk\lambda}$ we arrive at
\begin{eqnarray}
H&=&E_J\left(1-\cos\phi\right)+
  \left(\frac{E_C}{2}+
  \sum_{\vk\lambda}\frac{\mu_{\vk\lambda}^2}
  {\hbar\omega_k}\right)N^2+N\sum_{\vk\lambda}\mu_{\vk\lambda}
 \left(a_{\vk\lambda}+a_{\vk\lambda}^\dagger\right)+
  \sum_{\vk\lambda}\hbar\omega_k a_{\vk\lambda}^\dagger
  a_{\vk\lambda}\quad ,\label{ee9b}
\end{eqnarray}
where $\mu_{\vk\lambda}\equiv (8\pi
e^2\hbar\omega_k)^{1/2}\Lambda_{\vk\lambda}$. The Hamiltonian
(\ref{ee9b}) is of the Caldeira--Leggett type with a periodic
potential.  However, there is a subtle but important difference. If, as
is standard, we interpret $\phi$ as the position operator and $N$ as
the momentum, then we must conclude that the EM field couples to the
momentum operator instead of to the position. This type of coupling is
often referred to as anomalous coupling \cite{legg84,cucc01}.

\subsection{QED spectral density}
\label{Sd} In order to obtain the spectral density $J_{\rm
EM}(\omega)$, defined as
\begin{equation} \label{def-J-EM}
J_{\rm EM}(\omega)\equiv(2/\hbar^2 ) \sum_{\vk\lambda}
\mu_{\vk\lambda}^2\delta(\omega-\omega_k)
\end{equation}
one must in principle solve for the normal modes
$A_{\vk\lambda}(\vecr)$, which are the solutions of a complicated
boundary value problem. For simplicity, we approximate the EM normal
modes as those of free space with appropriate cutoffs which account for
the real geometry. We expect this to be a good approximation for
point-contact Josephson links while it will provide only a qualitative
description in the case of parallel plates separated by a dielectric.
To obtain $J_{\rm EM}(\omega)$ we substitute the sum by an integral,
$\sum_{\vk\lambda}\longrightarrow (V/8\pi^3)\sum_\lambda\int k^2\diff
k\;d(\cos\theta_k)\;d\phi_k$ and take advantage of the transversality
of the electromagnetic field $\vec{{\bf \varepsilon}}_\lambda\cdot{\bf
e}_k=0$. We choose $\vec{{\bf \varepsilon}}_1={\bf e}_{\theta}$ and
$\vec{{\bf \varepsilon}}_2={\bf e}_{\phi}$. With ${\bf
e}_{\phi}\cdot{\bf e}_z=0,{\bf e}_{\theta}\cdot{\bf e}_z=-\sin\theta_k$
and ${\bf e}_{k}\cdot{\bf e}_z=\cos\theta_k$ we arrive at
\begin{eqnarray}
J_{\rm EM}(\omega)&=&\frac{\alpha}{\pi^2}\;\omega
          \int_0^1 dy\left(\frac{1}{y^2}-1\right)
          \sin^2\left(\frac{\omega d}{c}y\right)\nonumber\\
          &=&
          \frac{\alpha\omega}{\pi^2}
          \left[\frac{\omega d}{c} \,{\rm Si}
          \left(\frac{2\omega d}{c}\right)
          -\sin^2\left(\frac{2\omega d}{c}\right)
          \frac{c}{4\omega d}\sin\left(\frac{2\omega d}{c}\right)
          -\frac{1}{2}\right]\, .
\label{ee7}
\end{eqnarray}
$J_{\rm EM}(\omega)$ has the cubic infrared behavior predicted by the
electric dipole approximation \cite{zapa97}. In the ultraviolet regime
it diverges quadratically. Specifically, we have
\begin{eqnarray}
J_{\rm EM}(\omega)&=&\frac{8\alpha d^2}{3\pi c^2}\omega^3
                         \quad {\rm for}\quad
                        \frac{\omega d}{c}\ll 1\ ,\nonumber\\
J_{\rm EM}(\omega)&=&\frac{2\alpha d}{c}\,\omega^2
                         \quad
                         {\rm for}\quad
                        \frac{\omega d}{c}\gg 1 \ .
\label{ee7aa}
\end{eqnarray}
Despite being a sum of oscillating functions, $J(\omega)$ is
monotonically increasing. The crossover from cubic to quadratic
behavior takes place on the scale of $c/d$. Finally, as already
mentioned above, the reduction of the two dimensional sine--Gordon
model to Eq.~(\ref{ee0}) requires a constant ${\bf A}_\parallel$. This
is taken into account by an ultraviolet cutoff function
$g(\omega/\omega_c)$. Any function $g$ with $g(0)=1$ whose modulus
vanishes at infinity faster than any power-law decay serves as a cutoff
function. One can choose also an algebraic cutoff function provided
that $\lim_{\omega\rightarrow\infty}J_{\rm EM}(\omega)=0$. We take the
cutoff frequency $\omega_c= c/\sqrt{A}$, where $A$ is the surface of
the electrodes. For point contact structures one usually has
$\sqrt{A}\gtrsim d$. Therefore, in order to obtain rough quantitative
estimates one may use the simple form of $J_{\rm EM}(\omega)$ in the
infrared limit given in Eq.~(\ref{ee7aa}) multiplied by the cutoff
function. We express all lengths of the structure relative to the
cutoff frequency $\omega_c\approx c/\sqrt{A}$. This leads to the
introduction of the aspect ratio $f=d/\sqrt{A}$. In a point-contact
experiment, we usually have $f\lesssim 1$. Now the spectral density
reads
\begin{equation}
J_{\rm EM} (\omega)=
 \gamma\,\frac{\omega^3}{\omega_c^2}\,g
 \left(\frac{\omega}{\omega_c}\right) \, ,
\label{ee7a}
\end{equation}
where the dimensionless coupling constant
\begin{equation}
\gamma\equiv \frac{8\alpha f^2}{3\pi}
\end{equation}
has been introduced, which hereafter will be treated as independent of
the cutoff. Eq.~(\ref{ee7a}) is the spectral function we will use in
the remainder of this work. Strictly speaking the cutoff
$\omega_c=c/\sqrt{A}$ only applies to $\vk_{\parallel}$. In the
direction perpendicular to the junction the cutoff length is much
smaller and of the order of the Thomas--Fermi screening length
$\lambda_{\rm TF}\simeq 10$\AA. However, since the perpendicular
component of the mode wave vector does not couple to $\phi$, we may
assign to $\int \diff k_z$ the cutoff which we wish. In particular, we
choose an isotropic cutoff for convenience.

\subsection{Equilibrium correlation functions}
\label{EQ} The Hamiltonian (\ref{ee9b}) is still difficult to handle
due to the nonlinearity of the $\cos \phi$ term. It has been studied
extensively in its different aspects and with a variety of different
methods \cite{schfgl83}. In our present study we are interested in the
equilibrium dynamics when $k_BT,E_C\ll E_J$. Then $\langle \phi^2
\rangle \ll 1$ and we can reasonably approximate $1-\cos\phi \simeq
\phi^2/2$. The problem reduces to Ullersma´s model for a damped
harmonic oscillator \cite{ulle66,haak85}. For thermal initial
conditions the symmetrized autocorrelation function of the particle
number $C^{(+)}_{\rm NN}(t)=\frac{1}{2}\langle[N(t)N(0)]_{+}\rangle$
and of the phase difference
$C^{(+)}_{\phi\phi}(t)=\frac{1}{2}\langle[\phi(t)\phi(0)]_{+}\rangle$
can be expressed in terms of the susceptibility \cite{weis99}
$\chi(\omega)=\chi^\prime+i\chi^{\prime\prime}$.
\begin{eqnarray}
C^{(+)}_{\rm NN}(t)&=&\frac{E_J}{2\pi\hbar} \int_{-\infty}^{\infty}
                  \chi^{\prime\prime}(\omega)\cos(\omega t)
                  \coth\left(\frac{\hbar\beta\omega}{2}\right)\,
                  \diff\omega \nonumber\\
C^{(+)}_{\phi\phi}(t)&=&\frac{\hbar}{2\pi E_J}
                        \int_{-\infty}^{\infty}\omega^2
                  \chi^{\prime\prime}(\omega)\cos(\omega t)
                  \coth\left(\frac{\hbar\beta\omega}{2}\right)
                  \,\diff\omega \nonumber\\
\chi^{-1}(\omega)&\equiv& \omega_{\rm
JP}^2-\omega^2-(E_J/\hbar)\widetilde{J}_{\rm EM}(\omega)\ ,
\label{EQ3}
\end{eqnarray}
In the last line of Eq.~(\ref{EQ3}) we have introduced the function
\begin{eqnarray}
\widetilde{f}(\omega)& \equiv & \omega^2\int_0^\infty
                      \diff\omega^\prime\frac{f(\omega^\prime)}
  {\omega^\prime\left({\omega^\prime}^2-\omega^2-i0^+\right)}
                     \nonumber\\
                     &\equiv& \omega^2P \int_0^\infty
                     \frac{f(\omega^\prime)}
                      {\omega^\prime\left({\omega^\prime}^2-\omega^2
                      \right)}\diff\omega^\prime
                      +i\frac{\pi}{2} f(|\omega|)\ ,
\label{RSt}
\end{eqnarray}
which is defined for any sufficiently well behaved function
$f(\omega)$. We note that $\widetilde{f}(\omega)$ is always symmetric,
although $f(\omega)$ is not necessarily so. One can consider
$\widetilde{f}(\omega)$ as a kind of symmetrized Riemann transform of
$f(\omega)$.

To evaluate the quantities in Eq. (\ref{EQ3}) we employ the cubic
spectral density given in Eq.~(\ref{ee7a}), which scales with
$\omega_c^{-2}$. This property guarantees finite results in the
Markovian limit $\omega_c\rightarrow\infty$. By definition, the cutoff
frequency is much larger than $E_J$ and the Josephson--plasmon
frequency $\omega_{\rm JP}$. Then, to leading order in $\omega_c^{-1}$
the contribution of the principal value is just $N_c\omega_c$, where
$N_c\equiv\int_0^\infty g(x)dx$. Below we take $g(x)=\Theta(1-x)$,
which implies $N_c=1$. At zero temperature, the integrals in
Eq.~(\ref{EQ3}) can be expressed in terms of the roots of the equation
\begin{equation}
-i\omega^3- \left(\frac{2\hbar\omega_c^2}{\pi\gamma E_J}+
\frac{2\omega_c}{\pi}\right)\omega^2 +\frac{2\hbar}{\pi\gamma
E_J}\omega_c^2\omega_{\rm JP}^2=0 \ . \label{EQ8a1}
\end{equation}
The polynomial has one purely imaginary root $i\lambda$ and a pair of
roots $iz,iz^*$. We obtain
\begin{eqnarray}
C_{\rm NN}^{(+)}(0)&=&\frac{1}{\pi^2\gamma}\frac{\omega_c^2}
                  {\left(\lambda-z\right)\left(\lambda-z^*\right)}
                  \left[{\rm Re}\left(\ln\frac{\omega_c^2+z^2}{z^2}-
                  \ln\frac{\omega_c^2+\lambda^2}{\lambda^2}\right)
                  -\frac{{\rm Re}\left(z-\lambda\right)}
                  {{\rm Im}\left(z\right)}
                  {\rm Im}\left(
                  \ln\frac{\omega_c^2+z^2}{z^2}\right)\right]\\
C_{\phi\phi}^{(+)}(0)&=&\frac{\hbar^2}{\pi^2E_J^2\gamma}
                  \frac{\omega_c^2}
                  {\left(\lambda-z\right)\left(\lambda-z^*\right)}
                 \left[{\rm Re}\left(\lambda^2
                 \ln\frac{\omega_c^2+\lambda^2}{\lambda^2}
                  -z^2\ln\frac{\omega_c^2+z^2}{z^2}\right)-
                  \frac{{\rm Re}\left(z-\lambda\right)}{{\rm Im}
                  \left(z\right)}
                  {\rm Im}\left(z^2\ln\frac{\omega_c^2+z^2}{z^2}
                  \right)\right] \ .
\label{EQ8a}
\end{eqnarray}
One calculates the parameters $\lambda$ and $z=z_1+iz_2$ perturbatively
in powers of $E_J/\omega_c$ and $\omega_{\rm JP}/\omega_c$.
\begin{eqnarray}
\lambda&=&\frac{2\hbar}{\pi E_J\gamma}\omega_c^2+\frac{2}{\pi}\omega_c+
          \frac{\pi E_J\gamma\omega_{\rm JP}^2}{2\hbar}
          \frac{1}{\omega_c^2}+
          {O}\left(\omega_c^{-3}\right)\nonumber\\
z_1&=&-\frac{\pi E_J\gamma\omega_{\rm
JP}^2}{4\hbar}\frac{1}{\omega_c^2}+
          {O}\left(\omega_c^{-3}\right) \\
z_2&=&\omega_{\rm JP}-\frac{E_J\gamma\omega_{\rm
JP}}{2\hbar}\frac{1}{\omega_c}+
         \frac{3E_J^2\gamma^2\omega_{\rm JP}}{8\hbar^2}
         \frac{1}{\omega_c^2}
         +{O}\left(\omega_c^{-3}\right)\ .\nonumber \label{EQ9}
\end{eqnarray}
The Josephson plasmon frequency acquires an imaginary part ($z_1$) only
to second order in $\omega_c^{-1}$. On the other hand, $z_2=-{\rm
Re}(iz)$ decreases by an amount $E_J\gamma\omega_{\rm
JP}N_c/2\hbar\omega_c$, which may be viewed as a {\em macroscopic Lamb
shift}. Unlike in the ohmic case, a crossover to overdamped
oscillations is not possible if $\omega_c$ remains large enough.
Expanding Eq.~(\ref{EQ8a}) in powers of $\omega_c^{-1}$ one obtains
\begin{eqnarray}
\langle\phi^2\rangle&=&\frac{1}{2}\sqrt{\frac{E_C}{E_J}}
                \left[1+\frac{E_J\gamma}{2\hbar\omega_{\rm JP}}
                +\left(\frac{E_J\gamma}{\hbar\omega_{\rm JP}}
                +\frac{3}{2}\right)\frac{E_J\gamma}{\hbar\omega_{c}}
                +{O}\left(\omega_c^{-2}\right)\right]\nonumber \\
\langle N^2\rangle&=&\frac{1}{2}\sqrt{\frac{E_J}{E_C}}
             \left[1-\frac{E_J\gamma}{2\hbar\omega_c}
             +{O}\left(\omega_c^{-2}\right)\right]\ .
\label{EQ9a}
\end{eqnarray}
In the limit $\gamma\to 0$ the susceptibility of the unperturbed
harmonic oscillator is recovered and we obtain the well-known results
for $\langle N^2\rangle$ and $\langle \phi^2\rangle$. The sign of the
correction reflects that the QED field tends to measure $N$ and the
expense of increasing the uncertainty in $\phi$.

For finite temperatures it is convenient to introduce an algebraic
cutoff function. We choose $g(x)=(1+x^4)^{-1}$. With this choice the
number of zeros of $\chi^{-1}(\omega)$ is finite and we are in a
position to evaluate both integrals in Eq.~(\ref{EQ3}) exactly. We
obtain the following low temperature dependence.
\begin{eqnarray}
\langle N^2\rangle&=&
           \langle N^2\rangle_{T=0}+
           \frac{\pi^4}{30}
           \frac{E_J\gamma}{\hbar^4E_C\omega_{\rm JP}^2\omega_c^2}
           (k_BT)^4
           +{O}(T^6)\nonumber\\
\langle\phi^2\rangle&=&
                 \langle\phi^2\rangle_{T=0}+
                 \frac{4\pi^4}{63}
                 \frac{\gamma}{\hbar^6\omega_{\rm JP}^4\omega_c^2}
                 (k_BT)^6
                 +{O}(T^8)\ .
 \label{EQ10}
\end{eqnarray}
Note that for $\omega_c\rightarrow\infty$ there are no finite $T$
corrections. A sketch of the derivation of these results is given in
Appendix \ref{ApD}.

The results in Eqs.~(\ref{EQ9a}) and (\ref{EQ10}) deserve some
discussion. First we concentrate on the mean square of the phase
difference. One can write the leading contribution to $\langle
\phi^2\rangle $ in Eq.~(\ref{EQ9a}) in the following way
\begin{equation}
\langle \phi^2\rangle
=\frac{1}{2}\sqrt{\frac{E_C}{E_J}}+\frac{2\alpha}{3\pi}f^2 \
.\label{EQ11}
\end{equation}
This equation is one of the central results of this work. It clearly
distinguishes between the uncertainty of the phase difference due to
the longitudinal electromagnetic field and that due to the transverse
electromagnetic field. The QED correction depends on the details of the
junction only through the aspect ratio $f$. Junction with $f$ close to
unity are possible in the case of point contacts. However, the
universal constant $2\alpha/3\pi\simeq 1.5 \times 10^{-3}$ renders the
influence of the transverse field a tiny effect. The contribution of
the longitudinal field contains several quantities which are
experimentally relevant. First, we have the critical current in
$E_J=\hbar I_c/2e$. It may be approximated by the formula of Ambegaokar
and Baratoff \cite{ambe63}
\begin{equation}
I_c(T)=\frac{\pi\Delta(T)}{2eR}\tanh\frac{\Delta(T)}{2k_BT}
\label{EQ12}
\end{equation}
at zero temperature. It may be viewed as decreased by the QED field
through and effective Debye-Waller factor. Second, we have the
capacitance in $E_C=4e^2/C$.

The case of a plate capacitance is characterized by $\sqrt{A}\gg d
>k_F^{-1}$. This limit corresponds to an aspect ratio $f \ll 1$.
The capacitance can be approximated by $C=\epsilon A/4\pi d$. The
normal state resistance of a tunneling junction is given by
$R^{-1}=e^2N(E_F)Ak_F^2 |T|^2/2$, where $N(E_F)$ is the density of
states per spin and $|T|^2$ is the average transmission probability for
electrons at the Fermi surface. For a rectangular barrier of height not
much greater than $E_F$, one may approximate
\begin{equation}
        |T|^2 \approx a \e^{-bk_F d'}\ ,
\label{EQ13}
\end{equation}
where $a$ and $b$ are numbers of order unity and $d'\lesssim d$ is the
barrier thickness. For qualitative estimates we may assume $d'=d$ which
underestimates slightly the QED effects. For simplicity, we also assume
$a\approx b \approx 1$ in the ensuing discussion. If we only include
the leading QED correction, Eq. (\ref{EQ9a}) becomes
\begin{equation} \label{phi-2-plate}
\langle \phi^2\rangle =\frac{2}{3\pi
A_F}\left[12\left(\frac{\pi^5\xi_0}{\epsilon a_0}\right)^{1/2}
\sqrt{d_F} \e^{d_F/2} +\alpha d_F^2\right] \ ,
\end{equation}
where $d_F\equiv k_Fd$, $A_F \equiv Ak_F^2$, $a_0$ is the Bohr radius,
and $\xi_0=\hbar v_F/\pi\Delta$ is the zero temperature coherence
length. Since typically $\xi_0\sim 2000\, a_0$, the ratio between the
two terms within square brackets is $\sim (150/\alpha)
d_F^{-3/2}\exp(d_F/2)$, which takes a minimum value of $\sim 10^4$ at
$d_F=3$. We conclude that, for plate capacitances and within the
approximation of free space photons, EM vacuum fluctuations contribute
negligibly to the quantum phase spread.

The situation is not much better if one considers the case $d \ll
\sqrt{A}$, as might correspond to e.g. a tunneling point contact. The
capacitance is then approximated as $C\sim \sqrt{A}$. Eq.
(\ref{phi-2-plate}) is replaced by
\begin{equation} \label{phi-2-pc}
\langle \phi^2\rangle =\frac{2}{3\pi
A_F}\left[12\left(\frac{\pi^5\xi_0}{\epsilon a_0}\right)^{1/2}
A_F^{1/4} \e^{d_F/2} +\alpha d_F^2\right] \ ,
\end{equation}
For the QED term to be comparable to the charging contribution, one
would need unrealistically small areas which would effectively suppress
the supercurrent.

The case of a planar junction which has been considered in the first
place presents the unfortunate property that both the longitudinal and
transverse contributions scale identically with the junction area. That
makes it difficult to overturn the natural smallness of the QED
correction. We note however that we have considered the spectrum of
photons in free space, an approximation that was admitted to be less
adequate for the case of broad Josephson junctions. In fact, the
structure of Eq. (\ref{EQ3}) suggests the possibility of designing
devices where the coupling of the phase to the photon field could be
amplified.

\section{Coupling to the electronic bath}
\label{HB}
\subsection{Effective Hamiltonian}
\label{effH} A complete treatment of the macroscopic phase dynamics in
Josephson junctions must include the effect of quasiparticles and
Cooper pairs. It was shown in Ref.~ \cite{ecke82} within a
path-integral language that the electronic degrees of freedom can be
taken into account by a coupling to two independent oscillator baths.
Starting from a microscopic tunneling Hamiltonian, they derived an
effective Lagragian in which the phase difference couples to two heat
baths which we label with superindices $(\pm)$. Here we propose an
alternative Hamiltonian--equation--of--motion approach to derive the
coupling and spectra of these effective baths:
\begin{eqnarray}
H&=&E_J\left(1-\cos\phi\right)+\frac{E_C}{2}N^2
 +\sum_i \hbar \omega_i\left[{b_i^{(+)}}^\dagger+
 \frac{\lambda_i^{(+)}}{\hbar\omega_i}
 \cos\left(\frac{\phi}{2}\right)\right]\left[b_i^{(+)}+
 \frac{\lambda_i^{(+)}}{\hbar\omega_i}
        \cos\left(\frac{\phi}{2}\right)\right]\nonumber\\
        &&\qquad\qquad\qquad
        +\sum_i \hbar\omega_i\left[{b_i^{(-)}}^\dagger+
        \frac{\lambda_i^{(-)}}{\hbar\omega_i}
        \sin\left(\frac{\phi}{2}\right)\right]
        \left[b_i^{(-)}+\frac{\lambda_i^{(-)}}{\hbar\omega_i}
        \sin\left(\frac{\phi}{2}\right)\right]
        \ .
\label{EE1012}
\end{eqnarray}
The coupling constants $\lambda_i^{(\pm)}$ are defined through the
spectral densities they must yield. Specifically, the expression
\begin{equation} \label{def-J}
J^{(\pm)}(\omega)\equiv\hbar^{-2}
\sum_i{\lambda_i^{(\pm)}}^2\delta(\omega-\omega_i)
\end{equation}
must be equal to
\begin{eqnarray}
J^{(\pm)}(\omega)&=&\frac{|T|^2}{2\hbar^2}\int \diff\omega^\prime
                   N_{\rm qp}(\omega^\prime+\frac{\omega}{2})
                  N_{\rm qp}(\omega^\prime-\frac{\omega}{2})
                  \left(1\pm\frac{\Delta^2/\hbar^2}
                  {\omega^{\prime 2}-\omega^2/4}\right)\nonumber\\
              &&\qquad\qquad\qquad\qquad
               \left\{\tanh\left[\frac{\hbar\beta}{2}
               \left(\omega^\prime
                  +\frac{\omega}{2}\right)\right]-
                  \tanh\left[\frac{\hbar\beta}{2}\left(\omega^\prime
                  -\frac{\omega}{2}\right)\right]\right\}\ ,
\label{aes01}
\end{eqnarray}
where
\begin{equation} \label{qpdensity}
N_{\rm qp}(\omega)=N(E_F)\Theta\left(\hbar\omega-\Delta\right)
             \frac{\hbar\omega}{\sqrt{\hbar^2\omega^2-\Delta^2}}
\end{equation}
is the quasiparticle density of states \cite{baro82,tink96}.

We derive this Hamiltonian in subsection \ref{qpheatbath} avoiding the
path--integral formalism used in Ref.~ \cite{ecke82}. Our criterion
will be that the heat baths must yield the correct equations of motion
for the phase $\phi(t)$. The reader who is familiar with the subject or
is not interested in this point might skip the next subsection an take
the Hamiltonian~(\ref{EE1012}) for given.

\subsection{Derivation of the effective Hamiltonian
             including coupling to electronic degrees of freedom}
\label{qpheatbath} Our starting point is the standard tunneling
Hamiltonian for two coupled superconductors in the Bogoliubov--de
Gennes mean--field approximation
 \cite{ecke82,kett99,ambe63,jose62,cohe62}.
\begin{eqnarray}
H_{\rm BdG}&=&\int\int_{\stackrel{\scriptstyle\vecr_1,\vecr_1^\prime\in
V_L}{\vecr_2,\vecr_2^\prime\in V_R}}
        \diff\vecr_1 \diff\vecr_1^\prime \diff\vecr_2 \diff\vecr_2^\prime
        {\bf \Psi}^\dagger(\vecr_1,\vecr_2)\;{\bf \Omega}\;{\bf
        \Psi}(\vecr_1^\prime,\vecr_2^\prime)\nonumber\\
{\bf \Psi}^\dagger(\vecr_1,\vecr_2)&\equiv&\ \
        \left[\psi^\dagger_{\uparrow}(\vecr_1),\psi_{\downarrow}(\vecr_1),
        \psi^\dagger_{\uparrow}(\vecr_2),
        \psi_{\downarrow}(\vecr_2)\right]\ ,
\label{em3}
\end{eqnarray}
with the $4\times 4$ matrix
\begin{equation}
{\bf\Omega}=
\left(\matrix{H(\vecr_1,\vecr^\prime_1)&\Delta(\vecr_1,\vecr^\prime_1)&
                       T_{\vecr_1,\vecr_2^\prime}&0\cr
                     \Delta^*(\vecr_1,\vecr^\prime_1)&
                     -H^*(\vecr_1,\vecr^\prime_1)
                     &0&-T^*_{\vecr_1,\vecr_2^\prime}\cr
                     T^*_{\vecr_1^\prime,\vecr_2}&0&
                     H(\vecr_2,\vecr^\prime_2)&
                     \Delta(\vecr_2,\vecr^\prime_2)\cr
                     0&-T_{\vecr_1^\prime,\vecr_2}&
                     \Delta^*(\vecr_2,\vecr^\prime_2)&
                     -H^*(\vecr_2,\vecr^\prime_2)
                     }\right)\ .
\label{em4}
\end{equation}
The order parameters $\Delta(\vecr,\vecr^\prime)$ obeys the
self-consistency condition within the bulk of each superconductor. In
Eq.~(\ref{em3}) we suppressed the dynamics of the electromagnetic field
for momentary convenience. The operators $H(\vecr,\vecr^\prime)$ as
well as
$\Delta(\vecr,\vecr^\prime)=
|\Delta(\vecr,\vecr^\prime)|\exp(i\phi(\vecr,\vecr^\prime))$
are local in space. After a further gauge transformation
\begin{equation}
U=\exp \left\{\frac{i}{2}\int \diff\vecr\,
  {\bf \Psi}^\dagger(\vecr)\,\diag\left[\phi(\vecr_1),\phi(\vecr_1),
  \phi(\vecr_2),\phi(\vecr_2)\right]\,{\bf \Psi}(\vecr)\right\}\ , \label{aes1}
\end{equation}
and in the grand canonical ensemble, the entries of ${\bf\Omega}$ in
Eq.~(\ref{em4}) look explicitly as follows:
\begin{eqnarray}
H_{L,R}(\vecr,\vecr^\prime)&=&\left[-\frac{\hbar^2}{2m}{\bf \nabla}^2+
         \frac{i\hbar}{2}[{\bf \nabla},{\bf v}_{L,R}(\vecr)]_+
         -\frac{m}{2}{\bf v}_{L,R}^2(\vecr)- \mu_{L,R}\pm\frac{2e^2N}{C}\right]
   \delta(\vecr-\vecr^\prime)\\
\Delta_{L,R}(\vecr,\vecr^\prime)&=&|\Delta_{R,L}(\vecr)|
   \delta(\vecr-\vecr^\prime)\nonumber\\
T_{\vecr_1,\vecr_2}&=&\exp(i\phi(\vecr_1,\vecr_2)/2)
\widehat{T}_{\vecr_1,\vecr_2}\ . \label{aes1a}
\end{eqnarray}
We remember here that $N$ is the semi-difference of Cooper--pairs
between the left and the right electrodes, $N=(N_1-N_2)/2$. We have
written the tunneling part of the Hamiltonian $H_T$ in such a way that
the gauge dependence becomes explicit.
\begin{eqnarray}
T_{\vecr,\vecr^\prime}&=&\widehat{T}_{\vecr,\vecr^\prime}
                      \exp i\Lambda(\vecr,\vecr^\prime)\nonumber\\
\Lambda(\vecr,\vecr^\prime)&=&\frac{e}{\hbar c}
                    \int_{\vecr^\prime}^\vecr
                    {\bf A}({\bf r})\cdot \diff\vecr
                   =\ -\Lambda(\vecr^\prime,\vecr)\ .
\label{em1a}
\end{eqnarray}
$\widehat{T}_{\vecr,\vecr^\prime}$ is gauge invariant and, without loss
of generality, can be assumed to be real symmetric. Thus the
fermion--fields $\psi_\sigma$ are the only quantities in
Eq.~(\ref{em3}) which break local gauge invariance. The notation above
may look unnecessary, since in the standard descriptions of Josephson
junctions \cite{kett99} the tunneling is treated locally and thus the
gauge term in (\ref{em1a}) can be safely dropped. However for our
purposes it is convenient to keep the gauge invariant line integral in
the definition of the tunnel matrix element. The virtue of the
transformation (\ref{aes1}) is that it permits to elucidate the role of
the phase of the order parameter, which appears in the transformed
Hamiltonian only in a gauge invariant form, through either ${\bf
v}(\vecr)$ or $\phi(\vecr,\vecr^\prime)=\phi(\vecr)-
\phi(\vecr^\prime)+\Lambda(\vecr,\vecr^\prime)$.

In order to derive an effective theory for the phase one makes two
further approximations. First, one assumes for simplicity that the
tunneling matrix element is independent of the electron momenta, i.e.
one sets $\widehat{T}(\vecr,\vecr^\prime)=T\delta(\vecr_{\parallel}-
\vecr_{\parallel}^\prime)\delta(z)\delta(z^\prime)$
\cite{comm-elsa,prad03}. Second, one approximates the unperturbed parts
$H_{L,R}(\vecr)$ substituting ${\bf v}(\vecr)$ and $\Delta(\vecr)$ by
their mean field bulk values ${\bf v}(\vecr)=0, z\neq 0$,
$\Delta(\vecr)=\Delta$. This is a reasonable approximation in the
tunneling limit \cite{ecke82,ambe63,jose62}. Using the above
approximations we may write the tunneling Hamiltonian as
\begin{eqnarray}
H&=&(E_C/2)N^2+H_L+H_R +H_T\nonumber\\
H_{L(R)}&=&\sum_{l(r),\sigma}\varepsilon_{l(r)}
           c_{l(r)\sigma}^\dagger c_{l(r)\sigma}+\Delta \sum_{l(r)}
           \left(c_{l(r)\uparrow}^\dagger
           c_{-l(r)\downarrow}^\dagger+{\rm H.c.}\right)\nonumber\\
       &=&\sum_{l(r),\sigma}
          E_{l(r)}\gamma_{l(r),\sigma}^\dagger
          \gamma_{l(r),\sigma} \nonumber\\
H_T     &=&T\cos(\phi/2)\sum_{l,r,\sigma}
           \left(c_{l\sigma}^\dagger c_{r\sigma}+{\rm H.c.}\right)
            + iT\sin(\phi/2)\sum_{l,r,\sigma}
           \left(c_{l\sigma}^\dagger c_{r\sigma}-{\rm H.c.}\right) \ .
           \label{aes31}
\end{eqnarray}
In the third line we have used the definition of the Bogoliubov
quasiparticle operators $\gamma_{k,\uparrow}=
u_kc_{k\uparrow}-v_{k}c_{-k\downarrow}^\dagger$ and
$\gamma_{k,\downarrow}=
u_kc_{-k\downarrow}+v_{k}c_{-k\uparrow}^\dagger$.

Now we adopt a bosonization procedure whereby the Hamiltonian
(\ref{aes31}) is mapped onto an effective Hamiltonian of the form
\begin{equation}
H_{\rm eff}=\frac{E_C}{2}N^2+\cos(\phi/2)\sum_i\lambda_i^{(+)}
        (b_i^{(+)}+b_i^{(+)\dagger})
 +\sin(\phi/2)\sum_i\lambda_i^{(-)}(b_i^{(-)}+
            b_i^{(-)\dagger})+
        \sum_{(\pm),i}\omega_ib_i^{(\pm)\dagger}b_i^{(\pm)}\ ,
\label{aes0}
\end{equation}
where the $b_i^{(\pm)}$ are bosonic operators and the coupling
strengths $\lambda_i^{(\pm)}$ are yet undefined. To further specify the
Hamiltonian (\ref{aes0}) we use the following strategy: We derive the
Heisenberg equations for the canonical conjugate variables $N$ and
$\phi$ as well as for the electronic degrees of freedom of the
Hamiltonian (\ref{aes31}). Then the equations for the electronic
degrees of freedom are solved in perturbation theory. Upon elimination
of the electronic degrees of freedom one obtains two coupled
integro--differential quantum Langevin equations \cite{weis99} for the
Heisenberg operators $N$ and $\phi$. It turns out that, up to second
order in the hopping $T$, these equations have the same structure as
those obtained from Eq.~(\ref{aes0}). This allows us to adjust the
coupling parameters $\lambda_i^{(\pm)}$ in Eq.~(\ref{aes0}) in such a
way that the equations of motion for $N$ and $\phi$ become identical.
Below we implement this strategy.

From the Hamiltonian (\ref{aes31}) we obtain the Heisenberg equations
\begin{eqnarray}
\dot{\phi}(t)&=&\hbar^{-1}E_C N(t)\nonumber\\
\dot{N}(t)&=&\frac{T}{2i\hbar}\sum_{l,r,\sigma}c_{l\sigma}^\dagger(t)
                  c_{r\sigma}(t)\e^{i\phi(t)/2}+{\rm H.c.} \,\, .
\label{aes001}
\end{eqnarray}
To close these equations we evaluate the time evolution of the
electronic part perturbatively in $H_T$. To first order we obtain
\begin{eqnarray}
c_{l\sigma}^\dagger(t) c_{r\sigma}(t)&=&
       c_{l\sigma}^{I\dagger}(t) c_{r\sigma}^{I}(t)
       +\frac{T}{i\hbar}\int_{-\infty}^t\diff t^\prime\sum_{l,r,\sigma}
       [c_{l\sigma}^{I\dagger}(t) c_{r\sigma}^{I}(t),
       c_{r\sigma}^{I\dagger}(t^\prime)
            c_{l\sigma}^{I}(t^\prime)]_-\e^{-i\phi(t^\prime)/2}+
            \nonumber\\
   &&\qquad\qquad\qquad\qquad
     \frac{T}{i\hbar}\int_{-\infty}^t\diff t^\prime\sum_{l,r,\sigma}
       [c_{l\sigma}^{I\dagger}(t) c_{r\sigma}^{I}(t),
       c_{-l-\sigma}^{I\dagger}(t^\prime)
            c_{-r-\sigma}^{I}(t^\prime)]_-\e^{i\phi(t^\prime)/2}\ ,
            \label{aes002}
\end{eqnarray}
and proceed similarly with the other relevant fermionic operators. The
superscript $I$ denotes operators in the interaction picture.
Introducing Eq.~(\ref{aes002}) into Eq.~(\ref{aes001}) yields
\begin{eqnarray}
\frac{\hbar}{E_C}\ddot{\phi}(t)&=&
                 \hat{F}(t)\exp\left(i\frac{\phi(t)}{2}\right)
                -\int^t_{-\infty}\diff t^\prime\hat{\alpha}(t-t^\prime)
                \exp\left(i\frac{\phi(t)-\phi(t^\prime)}{2}\right)
                \nonumber\\&&\qquad\qquad\qquad
               -\int^t_{-\infty} \diff t^\prime\hat{\beta}(t-t^\prime)
               \exp\left(i\frac{\phi(t)+\phi(t^\prime)}{2}\right)
               +{\rm H.c.}
                  \,\, . \label{aes003}
\end{eqnarray}
Since the uncoupled Hamiltonians $H_{L,R}$ in Eq.~(\ref{aes31}) are
quadratic forms, the corresponding grand canonical ensemble is a
Gaussian ensemble. Therefore the effect of the operator
\begin{equation}
\hat{F}(t)=\frac{T}{2i\hbar}
            \sum_{l,r,\sigma}c_{l\sigma}^{I\dagger}(t)
            c_{r\sigma}^{I}(t)
\label{aes004} \end{equation} is determined by its second moment
$\langle \hat{F}(t)\hat{F}(t^\prime)\rangle_{\rm B}$, where subscript B
denotes average over bath coordinates. These quantities are related to
the thermal averages of the operator valued dissipation kernels
\begin{eqnarray}
\hat{\alpha}(t)&=&\frac{|T|^2}{2\hbar^2}\sum_{l,r,\sigma}
                [c_{l\sigma}^{I\dagger}(t)c_{r\sigma}^{I}(t),
                c_{r\sigma}^{I\dagger}(0) c_{l\sigma}^{I}(0)]_-\,
                \nonumber\\
\hat{\beta}(t)&=&\frac{|T|^2}{2\hbar^2}\sum_{l,r,\sigma}
               [c_{l\sigma}^{I\dagger}(t)c_{r\sigma}^{I}(t),
                c_{-l-\sigma}^{I\dagger}(0) c_{-r-\sigma}^{I}(0)]_- \
\label{aes51}
\end{eqnarray}
by a fluctuation--dissipation theorem. In order to derive a
differential equation for the the $N$--point function
$\langle\phi(t_1)\ldots\phi(t_N)\rangle_{\rm B}$ from the Heisenberg
equations Eq.~(\ref{aes003}), one must in principle evaluate averages
of the form
$\langle\hat{\alpha}(t_1)\ldots\hat{\alpha}(t_N)\rangle_{\rm B}$.
However, since all averages are made over a Gaussian ensemble, Wick's
theorem applies and they reduce to products of
$\langle\hat{\alpha}(t_i)\rangle_{\rm B}$ and
$\langle\hat{\beta}(t_i)\rangle_{\rm B}$ respectively. Therefore, in
order to completely determine the quantum Langevin equation
Eq.~(\ref{aes003}), it is sufficient to calculate the averages
$\langle\hat{\alpha}(t)\rangle_{\rm B}$ and
$\langle\hat{\beta}(t)\rangle_{\rm B}$. These quantities are calculated
in standard textbooks on superconductivity \cite{tink96,kett99}. The
outcome is the well known result that
$\langle\hat{\alpha}(t)\rangle_{\rm B}$ and
$\langle\hat{\beta}(t)\rangle_{\rm B}$ are the sine--transforms of the
sum and difference, respectively, of the two spectral functions
$J^{(\pm)}(\omega)$,
\begin{eqnarray}
\langle\hat{\alpha}(t)\rangle_{\rm B}=\alpha(t)&=&
           \frac{1}{2}\int_0^\infty
           \left[J^{(+)}(\omega)+J^{(-)}(\omega)\right]
                 \sin(\omega t)\diff\omega\nonumber\\
\langle\hat{\beta}(t)\rangle_{\rm B}=\beta(t)&=&
          \frac{1}{2}\int_0^\infty
          \left[J^{(+)}(\omega)-J^{(-)}(\omega)\right]
                 \sin(\omega t)\diff\omega\,\, ,
\label{aes52}
\end{eqnarray}
where $J^{(\pm)}(\omega)$ are the spectral functions defined in
Eq.~(\ref{aes01}).

We can also derive a quantum Langevin equation for $N$ and $\phi$ from
the bosonized Hamiltonian Eq.~(\ref{aes0}). Since the interaction here
is linear in the bosonic operators, the procedure is exact (by
contrast, fermionic correlators are computed up to second order in
$H_T$). The resulting Heisenberg equation is identical to that in
Eq.~(\ref{aes003}) \cite{comm-jsc,sanc94}. However, in this case the
response kernels are given by
\begin{eqnarray}
\alpha(t)&=&\frac{1}{2\hbar^2}\sum_i\sin(\omega_i t)
             \left({\lambda_i^{(+)}}^2+ {\lambda_i^{(-)}}^2\right)
             \nonumber\\
\beta(t)&=&\frac{1}{2\hbar^2}\sum_i\sin(\omega_i t)
        \left({\lambda_i^{(+)}}^2-{\lambda_i^{(-)}}^2\right) \ .
        \label{aes521}
\end{eqnarray} Since the coupling parameters $\lambda^{(\pm)}_i$ have
not yet been specified we are free to choose them in a way that adjusts
the dissipation kernels in Eqs.~(\ref{aes521}) to the kernels
(\ref{aes52}) derived from the microscopic theory. Thus we may conclude
that the two Hamiltonians (\ref{aes31}) and (\ref{aes0}), together with
the spectral density (\ref{aes01}) are equivalent to order ${O}(T^2)$,
at least as far as the dynamics of the phase is concerned. We stress
that, due to the Gaussian nature of the averages, this equivalence
holds not only for the dynamics of the mean $\langle\phi(t)\rangle$ and
$\langle N(t)\rangle$ but also for the dynamics of the higher
correlators.

In order to make contact with the Hamiltonian Eq.~(\ref{EE1012}) we add
and subtract in Eq.~(\ref{aes0}) the expression
\begin{eqnarray}
&&\sum_i\frac{{\lambda^{(+)}}^2}{\hbar\omega_i}
  \cos^2\left(\frac{\phi}{2}\right)+
  \sum_i\frac{{\lambda^{(-)}}^2}{\hbar\omega_i}
  \sin^2\left(\frac{\phi}{2}\right)\
     =\nonumber\\
&&\qquad\qquad\qquad \frac{1}{2}\sum_i\frac{{\lambda^{(+)}}^2-
 {\lambda^{(-)}}^2}{\hbar\omega_i}\cos\phi
 +\frac{1}{2}\sum_i\frac{{\lambda^{(+)}}^2+
 {\lambda^{(-)}}^2}{\hbar\omega_i} \ , \label{erg}
\end{eqnarray}
which allows us to complete the square. The $\phi$ dependent term in
Eq.~(\ref{erg}) can be calculated using the definitions of
$\lambda_i^{(\pm)}$ in terms of the spectral densities $J^{(\pm)}$
given in Eq.~(\ref{aes01}). We obtain the result of Ambegaokar and
Baratoff \cite{ambe63}, namely,
\begin{eqnarray}
\frac{1}{2} \sum_i\frac{{\lambda^{(+)}}^2-{\lambda^{(-)}}^2}{\omega_i}
\cos\phi &=& \frac{1}{2}\int_0^\infty \diff\omega
\frac{J^{(+)}-J^{(-)}}{\hbar\omega}\cos\phi \nonumber\\
&=& \
 \frac{\hbar\pi\Delta(T)}{4e^2R}
 \tanh\left(\frac{\beta\Delta(T)}{2}\right)
 \cos\phi \nonumber\\
&=&\ E_J\cos\phi\ . \label{ABf}
\end{eqnarray}
The second term in the r.h.s. of Eq.~(\ref{erg}) is divergent but,
fortunately, it is independent of $\phi$. Since we are allowed to add
an arbitrary constant to a Hamiltonian, we subtract this infinity and
keep, for convenience, only a constant $E_J$ from this second term.
This yields the Hamiltonian (\ref{EE1012}).

Of course the two Hamiltonians  (\ref{EE1012}) and (\ref{aes0}) are
identical, but the physics is more transparent in Eq.~(\ref{EE1012}).
It is the Hamiltonian of a particle moving in a cosine potential
coupled nonlinearly to two independent bosonic baths. Since the
interaction is in the form of a complete square, the coupling to the
heat bath does not change the potential landscape. This is a well known
fact in the theory of dissipative quantum systems \cite{cald83,weis99}.

\section{Simultaneous coupling to photons and quasiparticles}
\label{simult}

\subsection{Mixed coupling Hamiltonian}
\label{mixed}

\label{MCH} If we now introduce again the dynamics of the
electromagnetic field as explained in Sec.~\ref{SECEM1} we arrive at a
rather comprehensive description of the dynamics of the phase in a
Josephson--junction. It is instructive to cast
 by a further unitary transformation the resulting Hamiltonian into a
 highly suggestive form.
 Applying $U^\dagger H U$ with
\begin{equation}
U =\exp\left[\cos\left(\frac{\phi}{2}\right)
\sum_i\frac{\lambda^{(+)}_i}{\hbar\omega_i}
  \left(b_i^{(+)}-{b_i^{(+)}}^\dagger\right)
  +\sin\left(\frac{\phi}{2}\right)
    \sum_i\frac{\lambda_i^{(-)}}{\hbar\omega_i}
  \left(b_i^{(-)}-{b_i^{(-)}}^\dagger\right)\right]
\label{m1}
\end{equation}
onto $H$ in Eq.~(\ref{EE1012}) shifts the interaction of the
``electronic" bosons with $\phi$ into an apparently more complicated
interaction with $N$. However, we can write the transformed Hamiltonian
as
\begin{eqnarray}
H&=&\frac{E_C}{2}\left(N+\delta
N\right)^2+E_J\left[1-\cos\left(\phi+\delta\phi\right)\right]
  \nonumber\\
&& \qquad+\sum_i\hbar\omega_i{b_i^{(+)}}^\dagger b_i^{(+)}
+\sum_i\hbar\omega_i{b_i^{(-)}}^\dagger
b_i^{(-)}+\sum_{\vk\lambda}\hbar\omega_k a_{\vk\lambda}^\dagger
a_{\vk\lambda} \ . \label{m2}
\end{eqnarray}
This form is rather general and physically quite elucidating. We
observe that the position degree of freedom of a one body Hamiltonian
acquires a fluctuating part. So does its canonical momentum. These
fluctuating parts obtain their own dynamics as bosonic excitations. The
specific form of these fluctuating operators  has either to be derived
from a many--body Hamiltonian or it has to be modelled
phenomenologically. In our case the fluctuating term of the particle
number
\begin{equation}
\delta N=\frac{i}{2}\sin\left(\frac{\phi+\delta\phi}{2}\right)\sum_i
      \frac{\lambda_i^{(+)}}{\hbar\omega_i}
  \left(b_i^{(+)}-{b_i^{(+)}}^\dagger\right)
  -\frac{i}{2}\cos\left(\frac{\phi+\delta\phi}{2}\right)
  \sum_i\frac{\lambda_i^{(+)}}{\hbar\omega_i}
  \left(b_i^{(+)}-{b_i^{(+)}}^\dagger\right)
\label{m3}
\end{equation}
is due to the hopping of Cooper pairs and quasiparticles across the
junction. It is phase dependent and was derived from a microscopic
model. An interesting message from Eq. (\ref{m2}) is that the Coulomb
interaction does not distinguish between $N$ and $\delta N$. It is only
sensitive to the overall particle number difference. This may be viewed
as a form of minimal coupling, since it reflects the ``gauge
invariance" of the Coulomb term and is analogous to the criterion
whereby the coupling to the EM field is introduced by replacing $\phi$
by its gauge invariant form. The fluctuating part of the phase is
$\delta\phi=\Lambda_{12}$ [see Eq.~(\ref{ee1a})].

Despite the appealing compactness of Eq.~(\ref{m2}), for calculational
purposes it is more convenient to shift both $N$ and $\phi$ by their
fluctuating part $-\delta N$ and  $-\delta\phi$ through a unitary
transformation. We employ the inverse of (\ref{m1}) to shift $N$ and
$U=\exp(-iN\Lambda_{12})$ to shift $\phi$. The resulting Hamiltonian is
much closer to the standard Caldeira--Leggett form
\begin{eqnarray}
&&H=E_J\left(1-\cos\phi\right)+\frac{E_C}{2}N^2
 +\sum_i\hbar\omega_i\left[{b_i^{(+)}}^\dagger+
 \frac{\lambda_i^{(+)}}{\hbar\omega_i}
 \cos\left(\frac{\phi}{2}\right)\right]\left[b_i^{(+)}+
 \frac{\lambda_i^{(+)}}{\hbar\omega_i}
        \cos\left(\frac{\phi}{2}\right)\right]\nonumber\\
        &&\quad+\sum_i\hbar\omega_i\left[{b_i^{(-)}}^\dagger+
        \frac{\lambda_i^{(-)}}{\hbar\omega_i}
        \sin\left(\frac{\phi}{2}\right)\right]
        \left[b_i^{(-)}+\frac{\lambda_i^{(-)}}{\hbar\omega_i}
        \sin\left(\frac{\phi}{2}\right)\right]
        +\sum_{\vk\lambda}\hbar\omega_k
        \left(a_{\vk\lambda}^\dagger+
        \frac{\mu_{\vk\lambda}}{\hbar\omega_k}N\right)
        \left(a_{\vk\lambda}+
        \frac{\mu_{\vk\lambda}}{\hbar\omega_k}N\right), \label{EE10}
\end{eqnarray}
with the coupling constants $\lambda_i^{(\pm)}$ defined by
$J^{(\pm)}(\omega)$ through Eqs. (\ref{def-J}) and (\ref{aes01}) and
$\mu_{\vk\lambda}$ defined by $J_{\EM}(\omega)$ through Eqs.
(\ref{def-J-EM}) and (\ref{ee7a}). The nonlinear coupling of the phase
to two different baths is rather difficult to handle. Following Ref.
\cite{ecke82}, one considerable simplification is achieved by keeping
only a local contribution in the $\hat{\beta}$ kernel integral of
Eq.~(\ref{aes003}). This is a valid approximation if the phase varies
slowly on the time scale in which decays $\beta(t)$, which is set by
$\hbar/\Delta$. Yet another simplification is obtained if one keeps
only the first non--vanishing term of the exponential in the
$\hat{\alpha}$ kernel integral in Eq.~(\ref{aes003}). With these
approximations the nonlinear coupling of the phase to the electronic
degrees of freedom has been replaced by a linear coupling to only one
effective heat--bath of an ``almost" ohmic character. These are the
basic assumptions which underlie the RCSJ model. The limits of validity
of these approximations have been discussed in the literature
\cite{weis99,ecke82}. In the RCSJ limit, the Hamiltonian (\ref{EE10})
becomes
\begin{eqnarray}
H&=&E_J\left(1-\cos\phi\right)+\frac{E_C}{2}N^2
        +\phi\sum_i\lambda_i(b_i+b_i^{\dagger})
        +N\sum_{\vk\lambda}\mu_{\vk\lambda}
        (a_{\vk\lambda}+a_{\vk\lambda}^\dagger)\cr
 & &\qquad+\sum_i\hbar\omega_ib_i^\dagger b_i+
     \sum_{\vk\lambda}\hbar\omega_k
     a_{\vk\lambda}^\dagger a_{\vk\lambda}
     + N^2\sum_{\vk\lambda}\frac{\mu_k^2}{\hbar\omega_k}+
     \phi^2\sum_i\frac{\lambda_i^2}{\hbar\omega_i}\ .
\label{ee11}
\end{eqnarray}
The coupling parameters $\lambda_i$ are adjusted to satisfy $J_{\rm
qp}(\omega)=2\hbar^{-2} \sum_i\lambda_i^2\delta(\omega-\omega_i)$,
where
\begin{eqnarray}
J_{\rm qp}(\omega) &\equiv& J^{(+)}(\omega)+J^{(-)}(\omega)
         \nonumber \\
&=&\frac{\hbar}{4\pi Re^2}\int \diff\omega^\prime
          N_{\rm qp}\left(\omega^\prime+
          \frac{\omega}{2}\right) N_{\rm qp}
          \left(\omega^\prime-\frac{\omega}{2}\right)
          \left\{\tanh\left[\frac{\hbar\beta}{2}
          \left(\omega^\prime+\frac{\omega}{2}\right)\right]-
          \tanh\left[\frac{\hbar\beta}{2}\left(\omega^\prime
         -\frac{\omega}{2}\right)\right]\right\} \ ,
\label{ee14}
\end{eqnarray}
$R$ being the normal resistance of the junction. This means that, in
the RCSJ approximation, diffusive effects due to $\beta(t)$ are
suppressed.

In the high temperature limit, Eq.~(\ref{ee14}) yields the ohmic
spectral density for a normal conductor junction given in
Eq.~(\ref{ohmic}): $J_{\rm qp}(\omega)\simeq\omega/R_Q$, $R_Q\equiv
2\pi Re^2/\hbar$. This ultimately justifies the RSCJ model. A plot of
$J_{\rm qp}(\omega)$ for various temperatures has been given by Harris
\cite{harr}. One observes that, already for relatively high
temperatures, the spectral density deviates from strictly Ohmic
behavior. In particular for $k_BT\lesssim \Delta/\hbar$ the spectral
density is better approximated by the zero temperature spectral
density, which can be calculated analytically \cite{wert}, the most
prominent feature being the existence of a gap at $\omega=2\Delta$. It
is known that an infrared gap in the spectral function reduces the
dissipative effect of the bath. Therefore, in the RCSJ model the effect
of quasiparticle tunneling is overestimated.

A central physical feature is that the two baths in Eq.~(\ref{ee11})
have competing effects. On the one hand, the quasiparticle environment
tends to ``measure" $\phi$ (the position of the particle).  On the
other hand the environment ``measures'' $N$ (the momentum of the
particle). Of course, due to the uncertainty principle, the two baths
cannot be equally effective. In general, it is not clear a priori which
mechanism will dominate. In this case, however, both because QED is
weak and superohmic, one expects the quasiparticle measurement of the
phase to be dominant.

Apart from its relevance in the present context, the model which we
have constructed is worthwhile to be studied in its own right
\cite{legg84}. We address this subject in a more general context in
Ref.  \cite{kohl03}. Here, however, we restrict ourselves to the case
in which one bath is ohmic (quasiparticles) and the other one is cubic
(photons).

\subsection{Equilibrium correlations}
\label{CqpEM} Like in Sec.~\ref{EQ}, we use the harmonic oscillator
approximation for the nonlinear potential terms. In contrast to the
case of a harmonic oscillator coupled to a single bath, this model is
not exactly solvable for an arbitrary initial condition. Fortunately,
we are mostly interested in the equilibrium quantities $C_{\rm
NN}^{(+)}(t)$ and $C_{\phi\phi}^{(+)}(t)$.
\begin{eqnarray}
&&C_{\rm
\phi\phi}^{(+)}(t)=\frac{1}{2\pi}
                      \int_{-\infty}^{\infty}|\chi(\omega)|^2
                  \cos(\omega t)\coth(\hbar\beta\omega/2)\cr
                  &&\qquad
                  \left\{
                  {\rm Im}\widetilde{J}_{\rm qp}(\omega)
                  \left(
                  \frac{E_C^2}{\hbar^2}-2\frac{E_C}{\hbar}
                  {\rm Re}\widetilde{J}_{\rm EM}(\omega)+
                  |\widetilde{J}_{\rm EM}(\omega)|^2
                  \right)+
                  \omega^2{\rm Im}\widetilde{J}_{\rm EM}(\omega)
                  \right\}\diff\omega \ , \nonumber\\
&&C_{\rm
NN}^{(+)}(t)=\frac{1}{2\pi}\int_{-\infty}^{\infty}|\chi(\omega)|^2
                  \cos(\omega t)\coth(\hbar\beta\omega/2)\cr
                  &&\qquad
                  \left\{
                  {\rm Im}\widetilde{J}_{\rm EM}(\omega)
                  \left(
                  \frac{E_J^2}{\hbar^2}-2\frac{E_J}{\hbar}
                  {\rm Re}\widetilde{J}_{\rm
                  qp}(\omega)+|\widetilde{J}_{\rm qp}(\omega)|^2
                  \right)
                  +\omega^2{\rm Im}\widetilde{J}_{\rm qp}(\omega)
                  \right\}\diff\omega\ .
\label{ee12}
\end{eqnarray}
Here, we have introduced the generalized ``susceptibility"
\begin{equation}
\chi^{-1}(\omega)=\omega_{\rm JP}^2-\omega^2-\hbar^{-1}E_J
             \widetilde{J}_{\rm EM}(\omega)-
             \hbar^{-1}E_C\widetilde{J}_{\rm qp}(\omega)+
              \widetilde{J}_{\rm qp}(\omega)
              \widetilde{J}_{\rm EM}(\omega)
\label{ee13}
\end{equation}
Again, $\widetilde{f}(\omega)$ denotes the symmetrized Riemann
transform as defined in Eq.~(\ref{RSt}). To leading order in
$\omega_c^{-1}$, Eq. (\ref{ee13}) reproduces Eq. (\ref{int6}), which
was derived from a simplified discussion. Of course, these expressions
reduce to Eqs.~(\ref{EQ3}) for $J_{\rm qp}(\omega)=0$. In
Eqs.~(\ref{ee12}) the spectral function $J_{\rm EM}(\omega)$ is defined
by Eq.~(\ref{ee7}), while for the quasiparticles we use the ohmic form
$J_{\rm qp}(\omega)=(\omega/R_Q) g(\omega/\Omega_c)$
keeping in mind that it overestimates dissipation at low temperatures.
Moreover we introduce formally an electronic cutoff $\Omega_c$, which
we may choose as the conduction band width. For the smallest geometries
available one has approximately $\Omega_c\gtrsim\omega_c$. In the
following we restrict ourselves to $T=0$. As pointed out in
Sec.~\ref{EQ}, this should be a good approximation at sufficiently low
temperatures $k_B T < \Delta \sim k_B T_c \ll \Omega_c $.

The evaluation of the integrals in Eq.~(\ref{ee12}) involves the
solution of a fifth order polynomial. Three of its solution scale as
$\sim \omega_c$ while the other two solutions shift the Josephson
plasmon frequency into the complex plane. Eventually the real part
disappears completely, leading to overdamped oscillations which are
characteristic of ohmic and subohmic spectral densities. For
$\langle\phi^2\rangle$ we find, to leading order in $\omega_c^{-1}$ and
$\Omega_c^{-1}$
\begin{eqnarray}
\langle\phi^2\rangle&=&\frac{1}{2}\,\sqrt{\frac{E_C}{E_J}}\,
        \frac{\ln[(\kappa+\sqrt{\kappa^2-1})/
        (\kappa-\sqrt{\kappa^2-1})]}
        {\pi\sqrt{\kappa^2-1}}\nonumber\\
        &&\qquad+f_\phi(\lambda_0,\lambda_1,\lambda_2)+
        {O}\left(\omega_c^{-1},\Omega_c^{-1}\right)
        \nonumber\\
f_\phi(\lambda_0,\lambda_1,\lambda_2)&=&
         \frac{2R_Q^2}{\pi^3\gamma^2}\,
         \frac{\gamma}{\Delta_3({\bf \lambda}^2)}
         \left|\matrix{g_\phi(\lambda_0)&\lambda_0^2&1\cr
                       \vdots&\vdots&\vdots}\right| \nonumber\\
g_\phi(\lambda_i)&=&\left(\frac{\pi\gamma}{R_Q}+
       \frac{\pi}{2}-\frac{\pi^3\gamma\lambda_i^2}
                            {8R_Q}
       \right)\ln\left(\frac{1+\lambda_i^2}{\lambda_i^2}\right)
                     \quad i=1\ldots3 \ ,
\label{ee16}
\end{eqnarray}
where we have used the notation $\kappa=(\pi/4R_Q)\sqrt{E_C/E_J}$.
Moreover, we have introduced Vandermonde's determinant $\Delta_N({\bf
r})\equiv\prod_{i<j}(x_i-x_j), \,\, i,j=1,..., N$. The dimensionless
quantities $\lambda_0,\lambda_1,\lambda_2$ are the coefficients of the
leading order term in $\omega_c$ of the three solutions of the fifth
order polynomial which scale with $\omega_c$. They are given by the
solution of the third order polynomial
\begin{equation}
\lambda^3+\left(\frac{2}{\pi}-\frac{\pi}{2}\right)\lambda^2-
   \frac{2}{\pi}\lambda+\frac{2R_Q}{\pi\gamma}=0 \ .
\label{ee17}
\end{equation}
Remarkably, the roots $\lambda_i$ only depends on the ratio
$R_Q/\gamma$. The expression for $\langle\phi^2\rangle$ in the first
line of Eq.~(\ref{ee16}) consists of two contributions. The first term
is recognized as the mean square of the position operator of a harmonic
oscillator with ohmic damping \cite{weis99}. It is the leading order
contribution in $\Omega_c$ of quasiparticle tunneling. The second term,
$f_\phi$, stems from the electromagnetic environment. Obviously the
exact analytic form of $f_\phi$ stated in Eq.~(\ref{ee16}) is not too
illuminating. To get further insight we consider the limiting cases
$R_Q/\gamma\ll 1$ and $R_Q/\gamma\gg 1$, which allow for an expansion
of the logarithm in $g_\phi$ and a perturbative solution of
Eq.~(\ref{ee17}). For $R_Q/\gamma\gg 1$ we obtain for $f_\phi$
\begin{equation}
f_\phi=\frac{\gamma}{4}+\frac{\gamma^2}{4R_Q}
       \left(1+\frac{\pi^2}{4}\right)
        +{O}\left(\frac{\gamma^2}{R_Q^2}\right) \ .
\label{ee19}
\end{equation}
As we have seen above, the parameter of the coupling to the
electromagnetic field $\gamma=8\alpha f^2/3\pi$ is always small.
However, the dimensionless resistance can be either large or of order
unity depending on the experiment. Assuming $R_Q\gg 1$ or,
equivalently, $\kappa\ll 1$, we can also expand the first relation in
Eq.~(\ref{ee16}). We obtain, to first order in $R_Q^{-1}$,
\begin{equation}
\langle\phi^2\rangle=\frac{1}{2}\sqrt{\frac{E_C}{E_J}}-
        \frac{E_C}{4E_JR_Q}
     +\frac{\gamma}{4}+\frac{\gamma^2}{4R_Q}
     \left(1+\frac{\pi^2}{4}\right)\ . \label{ee20}
\end{equation}
This result illustrates nicely the competing character of the effects
of the quasiparticle bath on the one hand and the photonic bath on the
other hand. The second term in Eq.~(\ref{ee20}) is the dominant
quasiparticle  contribution for $\Omega_c \rightarrow \infty$. It leads
to a reduction of $\langle\phi^2\rangle$ or, equivalently, to an
enhancement of coherence. The third term in Eq.~(\ref{ee20}) is the QED
correction to first order in $\omega_c^{-1}$, which was already found
in Eqs.~(\ref{EQ9a}) and ~(\ref{EQ11}) of Sec.~\ref{EQ}. This term is
always positive. One can estimate the quasiparticle term by using again
the Ambegaokar-Baratoff formula (\ref{EQ12}) and the capacitance of a
plate capacitor. One finds that the QED contribution would dominate for
an unrealistic thickness $d>12\hbar c/\epsilon\Delta\approx 100 \mu$m.

For the sake of completeness we also discuss the opposite limit
$R_Q/\gamma\ll 1$. After some algebra, we obtain that the leading order
term of the ``mixing'' function $f_\phi$ is
\begin{equation}
f_\phi=\frac{1}{4}R_Q
        \ln\left(\frac{\gamma}{R_Q}\right)\ .
\label{ee21}
\end{equation}
Expanding Eq.~(\ref{ee16}) for $R_Q\ll 1$ we obtain
\begin{equation}
\langle\phi^2\rangle=\frac{R_Q}{\pi^2}
   \left[4\ln\left(\frac{2}{R_Q}\right)+
   \frac{\pi^2}{2}\ln\left(\frac{\gamma}{R_Q}\right)\right] \ .
\label{ee22} \end{equation} The result for $\langle N^2\rangle$ is
formally similar to that of $\langle\phi^2\rangle$:
\begin{eqnarray}
\langle N^2\rangle&=&\frac{1}{2}\,\sqrt{\frac{E_J}{E_C}}\,
        \frac{\ln[(\kappa+\sqrt{\kappa^2-1})/
        (\kappa-\sqrt{\kappa^2-1})]}
        {\pi\sqrt{\kappa^2-1}}\nonumber\\
        &&+\frac{1}{2R_Q}\ln\left(\frac{\Omega_c}
        {\omega_{\rm JP}}\right)+
         f_N(\lambda_0,\lambda_1,\lambda_2)+
        {O}\left(\omega_c^{-1},\Omega_c^{-1}\right)
        \nonumber\\
f_N(\lambda_0,\lambda_1,\lambda_2)&=&
         \frac{2R_Q}{\pi^3\gamma^2}\,
         \frac{1}{\Delta_3(\lambda^2)}
         \left|\matrix{g_N(\lambda_0)&\lambda_0^2&1\cr
                       \vdots&\ddots&\vdots}\right|\nonumber\\
g_N(\lambda_i)&=&\left(\frac{\pi}{2\lambda_i^2}+
            \frac{\pi^3\gamma}{8R_Q}
            +\frac{\pi\gamma}{2R_Q}\lambda_i^2
            \right)\ln\left(\frac{1+\lambda_i^2}{\lambda_i^2}\right)
                     \quad i=1\ldots3 \ .
\label{ee23}
\end{eqnarray}
For small EM coupling ($\gamma/R_Q \ll 1$) the mixing function $f_N$
reduces to
\begin{equation}
f_N=\frac{\gamma}{4R_Q^2}\left(1+\frac{\pi^2}{4}\right)
        +{O}\left(\frac{\gamma^2}{R_Q^2}\right) \ .
\label{ee24} \end{equation}

We reproduce in Eq. (\ref{ee23}) the logarithmic dependence on the
quasiparticle cutoff frequency, a well known feature of the coupling to
an Ohmic heat bath \cite{weis99}.

\section{Summary}
\label{sum}

We have explored the role of the quantum electrodynamic field as a
possible agent of zero-point decoherence in some physical systems. We
have focussed on the fundamental problem of the dynamics of the
phase-number variable in a Josephson link. The transverse
electromagnetic field has been shown to couple to the relative Cooper
number through the electric dipole which it generates. The effect of
quasiparticles has been approximated with an effective oscillator bath
which couples to the phase. This poses a new quantum dissipation
problem of a type which has been little studied: that of a quantum
particle coupling through its position and momentum to two different
baths. The resulting dissipative dynamics of the macroscopic
phase-number variable has been investigated by means of a Hamiltonian
equation-of-motion approach. We have found that the r.m.s. values of
the phase and number contain interference contributions from the photon
and quasiparticle bath. The effect of the QED field has been found to
be quite small compared with charging effects due to Cooper pair and
quasiparticle fluctuations. The unfortunate fact that the contribution
of the longitudinal and transverse fields to the phase variance scale
identically with the junction area precludes a simple observation of
the effect of electromagnetic vacuum fluctuations in generic Josephson
devices. However, the sensitivity to the photon and quasiparticle
spectra which we have found suggests that the design of special
structures, where the effect of electrodynamic zero-point fluctuations
is amplified, cannot be ruled out.

\begin{acknowledgments} This work has been supported by Ministerio de Ciencia y
Tecnolog\'{\i}a (Spain) under Grants No. BFM2001-0172 and
MAT2002-0495-C02-01, and by the Ram\'on Areces Foundation. One of us
(H.K.) acknowledges financial support from the RTN Network of the
European Union under Grant No.~HPRN--CT--2000-00144.
\end{acknowledgments}

\begin{appendix}

\section{Derivation of Eqs.~(\ref{EQ10})}
\label{ApD} With the Drude type cutoff $g(x)=(1+x^4)^{-1}$ the inverse
susceptibility involves a fourth order polynomial in the numerator and
a second order polynomial in the denominator
\begin{eqnarray}
\chi^{-1}(\omega)&=&\frac{1}{\omega^2\pm i
\sqrt{2}\omega\omega_c-\omega_c^2}\left[\omega^4\pm
                  i\sqrt{2}\omega_c\omega^3\right.\nonumber\\
                  &&\left.+\left(\frac{\pi E_J}{2\sqrt{2}\hbar}
                  \gamma\omega_c
                  -\omega_{\rm JP}^2-\omega_c^2\right)\omega^2\mp
                  i\sqrt{2}\omega_{\rm JP}^2\omega_c\omega+
                  \omega_{\rm JP}^2\omega_c^2\right]
 \label{EQ5}
\end{eqnarray}
The roots of the fourth order polynomial come in two pairs of complex
numbers $i\lambda=i\lambda_1\pm \lambda_2$ and $iz=iz_1\pm z_2$. The
roots of the second order polynomial are given by
$\lambda_0=\omega_c\exp(\pm i\pi/4)$. One can evaluate Eqs.~(\ref{EQ3})
by a contour integral similarly to the Drude model \cite{ulle66,harr}.
However, here we will take advantage of the general expressions for the
mean squares $\langle N^2\rangle$ and $\langle\phi^2\rangle$ as
derivatives of the partition function of the harmonic oscillator
\cite{weis99}. We have
\begin{eqnarray}
\langle N^2 \rangle&=&-\frac{E_J}{2\beta\omega_{\rm JP}\hbar^2}
                              \frac{\diff}{\diff\omega_{\rm JP}}\ln Z(\beta)
                              \nonumber\\
\langle \phi^2 \rangle&=&-\frac{1}{2E_J\beta}
                              \left(\omega_{\rm JP}\frac{\diff}
                              {\diff\omega_{\rm JP}}
                              +2\gamma\frac{d}{d\gamma}\right)
                              \ln Z(\beta)\quad ,
\label{EQ6}
\end{eqnarray}
with the partition function of the damped harmonic oscillator
\begin{equation}
Z(\beta)=\frac{1}{\hbar\beta\omega_{\rm JP}}\prod_{n=1}^\infty
\nu_n^2\,\chi(|\nu_n|)\quad , \label{EQ7}
\end{equation}
We have used the notation $\nu_n=2\pi n/\hbar\beta$ for the Matsubara
frequencies.

At this point it proves useful to use an algebraic cutoff. That yields
a finite number of roots which satisfy a series of algebraic relations,
called Vieta relations \cite{abram72}. This allows us to express the
infinite product in Eq.~(\ref{EQ7}) compactly in terms of
$\Gamma$--functions
\begin{equation}
Z(\beta)=\frac{\hbar\beta\omega_{\rm JP}}{4\pi^2}
\frac{|\Gamma(\lambda/\nu)|^2|\Gamma(z/\nu)|^2}
      {|\Gamma(\lambda_0/\nu)|^2}
\label{apD0}
\end{equation}
The real and the imaginary parts of the roots $\lambda,z$  scale quite
differently with the cutoff frequency. They are approximately
\begin{eqnarray}
\lambda_1 & = &\frac{1}{\sqrt{2}}\omega_c+{O}(\omega_c^{-2})\cr
\lambda_2 &=&\frac{1}{\sqrt{2}}\omega_c-\frac{E_J\pi\gamma}{\hbar4}
             -\frac{\pi^2E_J^2}{16\sqrt{2}\hbar^2}
             \frac{\gamma^2}{\omega_c}
             +{O}(\omega_c^{-2})\cr
z_1      &  =&-\frac{\pi}{4}\frac{E_J\gamma\omega_{\rm
JP}^2}{\hbar\omega_c^2}+
              {O}(\omega_c^{-3})\cr
z_2      &  =&\omega_{\rm JP}+\frac{\pi}{4\sqrt{2}}
              \frac{E_J\gamma\omega_{\rm JP}}{\hbar\omega_c}+
              \frac{3\pi^2}{64}\frac{E_J^2\gamma^2\omega_{\rm JP}}
              {\hbar^2\omega_c^2}+{O}\left(\omega_c^{-3}\right)
\label{ApD1}
\end{eqnarray}
Using the asymptotic expansion of the $\Gamma$--function we get an
expansion of $Z(\beta)$  in powers of $k_BT$.
\begin{equation}
Z(\beta)=Z^{(0)}+\sum_{n=1}^\infty Z^{(n)}(\beta)\
              \left(k_B T\right)^n
\ . \label{ApD1a}
\end{equation}
In zeroth order we recover Eqs.~(\ref{EQ9a}) with $N_c=\pi/2\sqrt{2}$.
The next two orders in the expansion vanish identically due to the
Vieta relations. The first non--vanishing correction term reads
\begin{equation}
Z^{(3)}(\beta)=\frac{1}{360\omega_c^6\omega_{\rm JP}^6}
   \left[\left(\lambda^3+{\lambda^*}^3\right)|z|^6
   +\left(z^3+{z^*}^3\right)|\lambda|^6
   -\omega_{\rm JP}^6\left(\lambda_0^3+{\lambda_0^*}^3\right)\right]
   \quad .
\label{ApD2}
\end{equation}
Using Eq.~(\ref{EQ6}) and retaining only the highest order term in
$\omega_c$ we finally arrive at the result for $\langle N^2\rangle$
stated in Sec.~\ref{EQ}, with a quartic finite temperature correction.
Moreover we observe that for $\langle \phi^2\rangle$ the quartic
contribution vanishes as $\omega_c \rightarrow \infty$. Therefore, in
order to obtain the first finite temperature correction for $\langle
\phi^2\rangle$, we have to expand $Z(\beta)$ up to fifth order in
$\beta$. The result with a sixtic finite temperature correction is
given in Eq.~(\ref{EQ10}).
\end{appendix}


\begin{thebibliography}{99}
\bibitem{comm-elsa} A more physical local tunneling Hamiltonian,
where $\delta(z)$ is replaced by $\delta'(z)$, has been proposed in
Ref. \cite{prad03}. The change has consequences on the energy
dependence of the hopping matrix elements and thus on the average
transmission $|T|^2$,
which is correctly given by the choice $\delta'(z)$. However, this is
not a problem in the context of the Ambegaokar-Baratoff formula because
the product $I_c R$ is independent of the actual value of the average
transmission.
\bibitem{comm-jsc} The fact that the baths couple to
$\sin(\phi/2)$ and $\cos(\phi/2)$ eliminates unwanted dependence on the
initial conditions \cite{sanc94}.
\end{thebibliography}
\end{document}